\documentclass[twocolumn,showpacs,preprintnumbers,amsmath,amssymb]{revtex4}

\usepackage{graphicx,psfrag}
\usepackage{dcolumn}
\usepackage{bm}
\usepackage[T1]{fontenc}
\usepackage[latin1]{inputenc}
\usepackage{graphicx}
\usepackage{amssymb}

\newtheorem{theorem}{Theorem}
\newtheorem{corollary}{Corollary}
\newtheorem{lemma}{Lemma}
\newtheorem{remark}{Remark}

\begin{document}

\title{Improving the consensus performance via predictive mechanisms}
\author{Hai-Tao Zhang$^1$, Guy-Bart Stan$^1$, Michael ZhiQiang Chen$^{1,2,\dagger}$, Jan M. Maciejowski$^1$, and Tao Zhou$^{3,4}$}
\affiliation{$^1$Department of Engineering, University of Cambridge, Cambridge CB2 1PZ, U.K. \\
$^2$Department of Engineering, University of Leicester, Leicester
LE1 7RH, U.K.\\
$^3$Department of Modern Physics, University of Science and Technology of China, Hefei 230026, PR China\\
$^4$Department of Physics, University of Fribourg, Chemin du Muse,
CH-1700 Fribourg, Switzerland}

\begin{abstract}
  Considering some predictive mechanisms, we show that ultrafast
  average-consensus can be achieved in networks of interconnected
  agents. More specifically, by predicting the dynamics of the network
  several steps ahead and using this information in the design of the
  consensus protocol of each agent, drastic improvements can be
  achieved in terms of the speed of consensus convergence, without
  changing the topology of the network. Moreover, using these
  predictive mechanisms, the range of sampling periods leading to
  consensus convergence is greatly expanded compared with the routine
  consensus protocol. This study provides a mathematical basis for the
  idea that some predictive mechanisms exist in widely-spread
  biological swarms, flocks, and networks. From the industrial
  engineering point of view, inclusion of an efficient predictive
  mechanism allows for a significant increase in the speed of
  consensus convergence and also a reduction of the communication
  energy required to achieve a predefined consensus performance.

\end{abstract}

\maketitle

\section{Introduction}\label{sec:Introduction}

Over the last decade, scientists have been looking for some common,
possibly universal, features of the collective behaviors of animals
\cite{ma07}, bacteria \cite{bud95}, cells \cite{yo04}, molecular
motors \cite{as94}, as well as driven granular objects \cite{vi95}.
The collective motion of a group of autonomous agents (or particles)
is currently a subject of intensive research that has potential
applications in biology, physics and engineering. One of the most
remarkable characteristics of complex dynamical systems such as flocks
of birds, schools of fish, or swarms of locusts, is the emergence of a
state of collective orders in which the agents move in the same
direction, i.e. an ordered state \cite{vi95,gr04,al07}. This ordered
state seeking problem can be further generalized to a consensus
problem \cite{sa04, re07}, where a group of self-propelled agents
agree upon certain quantities of interest such as  attitude, position,
temperature, voltage, etc. Furthermore, solving consensus problems
using distributed computational methods has direct implications on
sensor network data fusion, load balancing, swarms/flocks, unmanned
air vehicles (UAVs), attitude alignment of satellite clusters,
congestion control of communication networks, multi-agent formation
control, and so on \cite{ak02,og04,ar02}.

Among the most important early works on consensus problems, Fiedler
\cite{fi73} showed that the second smallest eigenvalue $\lambda_2$,
namely the \textit{algebraic connectivity}, of the Laplacian matrix
$L$ associated with the graph defining the network topology is
directly related to the consensus speed of the network. It was also
shown that a network with high algebraic connectivity is robust to
both node-failures and edge-failures. In \cite{sa04}, Olfati-Saber and
Murray analyzed consensus problems in networks of agents with
continuous-time dynamics (basically integrators), switching topology
and time-delays, and proved that decreasing the largest eigenvalue
$\lambda_N$ of the Laplacian matrix $L$ improves the consensus
robustness of the network to time-delays. Consequently, the condition
number $\lambda_N/\lambda_2$ provides a measure of the consensus
performance of the considered network in the sense that the smaller
its value, the better the consensus performance (consensus speed and
robustness to time-delays). Moreover, in \cite{sa04}, the authors also
provided a feasible range of sampling rates leading to consensus in
the case of discrete-time dynamic networks. In this way, the
theoretical foundations of general consensus problems were
established.
To improve the speed of convergence towards consensus, they further
proposed a method based on the addition of a few long links to a
regular lattice, thus transforming it into a small-world network
\cite{sa05, st01}. In \cite{xi04}, Xiao and Boyd transformed the
fastest distributed linear averaging problem into a convex
optimization problem by considering a particular per-step convergence
optimization index. Additionally, they proved that, when the network
topology is symmetric, the problem of finding the fastest converging
linear iteration can be cast as a semidefinite programming problem,
and thus can be efficiently and globally solved. In \cite{ya06},
consensus problems in a heterogeneous influence network were
investigated by Yang at al. and it was shown that, by decreasing the
scaling exponent in the associated power-law distribution, the ability
of the network to reach direction consensus among its agents is
significantly enhanced due to the leading roles played by a few hub
agents.

In summary, most of the previous works achieved performance
improvements, such as increasing the consensus speed, improving the
robustness to nodes and edges failures, or improving the ability to
deal with time-delays, solely based on the currently available
information flows on the network. In these works, the computing
abilities of an agent of the network is fairly limited: the agent can
only observe the current behavior (state) of its neighbors and update
its state according to this observation. However, in natural
bio-groups, individuals generally possess some higher level of
predictive computing capabilities that they use for updating their
state. Experiments revealing the use of prediction mechanisms by
individuals of a network have been described in the literature
previously. In particular, as early as 1959, Woods \cite{wo59}
designed some bee swarms experiments and provided evidences for the
existence of certain predictive mechanism in bee swarms formation. In
1995, Montague {\it et al.} proposed simple hebbian learning rules to
explain the predictive mechanisms used by bees when foraging in
uncertain environments \cite{mo95}.  Apart from the investigation of
the predictive mechanisms used during swarming and foraging, several
researchers focused on the predictive functions of the optical and
acoustical apparatuses of individuals inside bio-groups, especially
the retina and cortex \cite{go03,su06,me07}. Based on intensive
experiments on the bio-eyesight systems, it was found that, when an
individual observer prepared to eye-follow a displacement of the
visual stimulus, the visual form of adaptation was transferred from
the current fixation to the future gaze position. These reported
investigations support the conjecture of the existence of some
predictive mechanisms inside many bio-groups.

\begin{figure}\label{fig: Bird_prediction}
  \centering
  \scalebox{0.4}[0.4]{\includegraphics{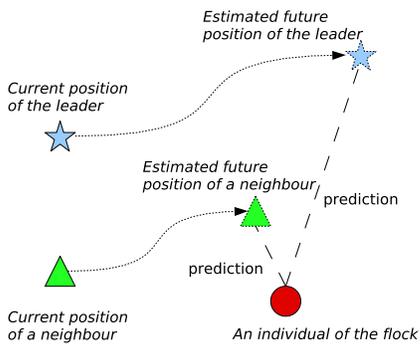}}
\caption{(Color online) Illustration of the predictive nature of
  flocks, swarms and networks.}
\end{figure}

In this paper, we show that, by introducing a predictive mechanism, it
is possible to either significantly enhance the consensus speed
obtainable under the constraint of a fixed amount of communication
energy or to decrease the communication energy required to ensure a
prescribed consensus speed. This observation allows us to infer that
prediction mechanisms may exist universally in many natural
bio-groups. A general physical picture illustrating this paradigm is
given in Fig.~1, which is interpreted as follows: in widely-spread
natural bio-groups composed of animals, bacteria, cells, etc., the
next-step behavioral decision of each individual/particle/agent is not
only based on the current available state information (position,
velocity, etc.) of the other (neighboring) agents inside the group,
but is also based on their predicted future states. More precisely,
bearing in mind a few past states of its leader and neighbors, an
agent can estimate their future states several steps ahead and then
make a decision on its own action. Estimation of these future states
by each agent can eliminate the requirement of intensive communication
among the agents so that the overall communication energy of the group
can be reduced effectively.

From an industrial application point of view, the phenomena and
mechanisms reported in this paper may be applicable in some relevant
prevailing engineering areas such as  autonomous robot formations, sensor
networks and UAVs \cite{ak02,og04,ar02}. Each agent in such a group
typically has limited power to send messages, and thus a larger
sampling period is generally desirable to save communication
energy. Since using a predictive mechanism can sharply expand the
range of feasible sampling periods, it should be useful for industrial
applications where overall communication energy is limited.

The rest of this paper is organized as follows. In
Section~\ref{Problem description}, the two main problems addressed in
this paper are formulated. In Section~\ref{Model predictive consensus
  protocol for all-to-all link networks}, the all-to-all link
\textit{model predictive control} (MPC) average-consensus protocol is
presented together with its convergence analysis and some statistical
simulation results. In Section~\ref{Model predictive consensus
  protocol for partial link networks}, the MPC consensus protocol is
generalized to partial link networks.  Additionally, conditions that
guarantee asymptotic convergence of the proposed protocol are
provided, and simulation results showing its main characteristics and
advantages are presented. Finally, conclusions are drawn in
Section~\ref{Conclusion}.

\section{Problem description}
\label{Problem description}

In order to discuss and illustrate the role of prediction mechanisms
in networks of interconnected agents, we consider one of the simplest
type of networks, i.e. networks of linearly interconnected
integrators. The interconnection structure of such networks can
typically be represented by a directed graph (called \textit{digraph}
for short). In this representation of the network, the $N$ nodes of
the digraph represent the $N$ agents of the network and a weighted
edge $a_{ij}$ from node $i$ to node $j$ indicates the existence of a
communication link from agent $i$ to agent $j$ in the network. The
digraph is denoted by $G=(\mathcal{V},\mathcal{E},A)$, where
$\mathcal{V}=\left\{ v_{1},\ldots,v_{N}\right\}$ is the set of nodes,
$\mathcal{E}\subset\mathcal{V}\times\mathcal{V}$ is the set of edges,
and $A$ is the associated weighted \textit{adjacency matrix},
i.e. $A=\left\{a_{ij}\right\}_{i,j=1,\ldots,N}\in \mathbb{R}^{N \times
  N}$, with nonnegative elements $a_{ij}$ which are zero when
there is no communication link from $i$ to $j$.  Furthermore, we
assume that there is no self-cycle, i.e. $a_{ii}=0$, $\forall
i=1,\ldots,N$.

Let $x_i(t)\in \mathbb{R}$ denote the state of node $i$, which might
represent physical quantities such as attitude, position, temperature,
voltage, etc. To obtain a full representation of the dynamic network
under consideration, we refer to it as $G_x=(G,x)$ where $x \in
\mathbb{R}^{N}$ denotes the network state and $G$ its
\textit{topology} (or information flow). Generally, we say that the
nodes of a network have reached \textit{consensus} if and only if
$x_i=x_j$ for all $i,j \in \mathcal{V}$. Whenever the nodes of a
network are all in agreement, their common value is called the
\textit{group decision value}. If the group decision value is
$\overline{x}(0)\triangleq 1/N(\sum_{i=1}^Nx_i(0))$, the network is
said to have reached the \textit{average-consensus}. In the rest of
the paper, for brevity, we denote average-consensus by consensus.

Agents with continuous-time models are typically described by the
integrator dynamics
\begin{equation}
  \label{eq: continuous model}
  \dot{x}_i(t)=u_i(t),
\end{equation}
while agents with discrete-time models are described by the difference
dynamics
\begin{equation}
  \label{eq: discrete model}
  x_i(k+1)=x_i(k)+\epsilon u_i(k),
\end{equation}
where $\epsilon$ denotes the sampling period or
step-size. Average-consensus is typically asymptotically reached using
the \textit{routine} consensus protocol
\begin{equation}
  \label{eq: routine control}
  u_i(t)=-\sum_{j=1}^{N}a_{ij}\Delta x_{i,j}(t),~~i,j=1,\ldots,N,
\end{equation}
where $\Delta x_{i,j}(t)\triangleq x_i(t)-x_j(t)$ denotes the state
difference between the $i^{th}$ and the $j^{th}$ agents of the
network. It has been proven in \cite{sa04} that, in a network $G_x$
with fixed topology and continuous-time dynamics determined by
(\ref{eq: continuous model}), the routine protocol (\ref{eq: routine
  control}) globally and asymptotically solves the consensus problem
if and only if $G$ is \textit{strongly connected} and
\textit{balanced}. The assumption of a strongly connected network
amounts to imposing that any two distinct nodes can be connected via a
path that follows the direction of the edges of the digraph. The
balanced network assumption corresponds to
$\sum_ja_{ij}=\sum_ja_{ji},~i=1,\ldots,N$ which is obviously more
general than the \textit{symmetric network} condition, in which
$a_{ij}=a_{ji},~i,j=1,\ldots,N$.


Considering the routine protocol (\ref{eq: routine control}), the
dynamics of a network of continuous-time integrator agents is defined
by:
\begin{equation}
  \label{eq: continuous-time network dynamics}
  \dot{x}(t)=-Lx(t),
\end{equation}
where $L=\left\{l_{ij}\right\}_{i,j=1,\ldots,N} \in \mathbb{R}^{N
  \times N}$ is called the \textit{graph Laplacian matrix} induced by
the topology $G$ and is defined as $l_{ii}=\sum_{l\neq i}^{N}a_{il}$,
$\forall i=1,\ldots,N$ and $l_{ij}=-a_{ij}$, $\forall i\neq j$, where
$a_{ij}$ denotes the $(i,j)$ element of the adjacency matrix
associated with $G$.  By construction, the Laplacian matrix has zero
row sum, i.e. $L\textit{\textbf{1}}=0$ with
$\textit{\textbf{1}}\triangleq [1,\ldots,1]^T_{N\times 1}$.


For agents with discrete-time dynamics (\ref{eq: discrete model}),
applying the discrete-time version of the routine consensus protocol
(\ref{eq: routine control}) yields the following discrete-time network
dynamics:
\begin{equation}
  \label{eq: discrete-time network dynamics}
  x(k+1)=P_\epsilon x(k)
\end{equation}
with $P_\epsilon=I_N-\epsilon L$ (see \cite{sa04}). Let $d_{\max}=
\max_{i}\left(l_{ii}\right)$ denote the maximum node out-degree of the
digraph $G$. As shown in \cite{sa04} and \cite{sa07}, if the network
is strongly connected and balanced, and the sampling period $\epsilon
\in (0,1/d_{\max})$, the routine consensus protocol (\ref{eq: routine
  control}) ensures global asymptotic convergence to consensus.

Some previous works were devoted to accelerating the speed of convergence
towards consensus (see for example \cite{sa04, sa07, sa05, xi04})
since high-speed consensus is obviously always desirable in
engineering practice. On the other hand, extension of the consensus
feasible range of the sampling period $\epsilon$ is also important as
it typically allows for a reduction in the required communication
energy and may provide an explanation for the fundamental mechanisms
used by bio-groups to exhibit collective behaviors. Accordingly, two
naturally-motivated problems are addressed in this paper:

\begin{itemize}
\item the increase of the average-consensus speed;
\item the extension of the feasible range of the sampling period.
\end{itemize}
The approach to be taken is based on the introduction of some
predictive mechanism.


For simplicity, we assume that all the networks considered in the rest
of the paper are strongly connected.

\section{Model predictive consensus protocol for all-to-all link
  networks}
\label{Model predictive consensus protocol for all-to-all link
  networks}

In this section, we first introduce an MPC algorithm to solve the
average-consensus problem for \textit{all-to-all link networks} (or
\textit{complete graphs} in which each pair of vertices is connected
by an edge). We then provide some theorems that support this
algorithm. Afterwards, we give some simulation results to illustrate
the feasibility and superiority of this MPC consensus algorithm.

\subsection{Algorithm}
In order to improve the consensus performances, we replace the routine
control protocol given in (\ref{eq: routine control}) by the following
MPC consensus protocol:

\begin{equation}\label{eq: MPC law}
  u_i(k)=\sum_{j=1}^{N}a_{ij}\Delta x_{i,j}(k)+v_i(k),
\end{equation}
where $v_i(k)$ is an additional term representing the MPC action, and
the state difference $\Delta x_{i,j}(k)\triangleq x_i(k)-x_j(k)$. With
this MPC protocol, the network dynamics are given by

\begin{equation}\label{eq: closed-loop system with MPC law}
  x(k+1)=P_\epsilon x(k)+v(k)
\end{equation}
with $v(k)=[v_1(k),\ldots,v_N(k)]^{T}$ representing the MPC decision
values for the $N$ nodes of $G$. The MPC element $v(k)$ will be
calculated by solving the optimization problem associated with a
specific moving-horizon optimization index function (see (\ref{eq:
  optimization index}), for example).

Using the consensus protocol (\ref{eq: MPC law}), the future network
state can be predicted based on the current state value $x(k)$ as
follows:

\[\begin{array}{c}
  x(k+2)=P_{\epsilon}^{2}x(k)+P_{\epsilon}v(k)+v(k+1),\\
  \vdots\\
  x(k+H_u)=P_{\epsilon}^{H_u}x(k)+\sum_{j=0}^{H_u-1}(P_{\epsilon}^{H_u-j-1}v(k+j)),\\
  x(k+H_u+1)=P_{\epsilon}^{H_u+1}x(k)+\sum_{j=0}^{H_u-2}(P_{\epsilon}^{H_u-j}v(k+j))\\
  +(P_{\epsilon}+I)v(k+H_u-1),\\
  \vdots\\
  x(k+H_p)=P_{\epsilon}^{H_p}x(k)+\sum_{j=0}^{H_u-2}(P_{\epsilon}^{H_p-j-1}v(k+j))\\
  +\sum_{j=0}^{H_p-H_u}P_{\epsilon}^jv(k+H_u-1).
\end{array}\]

Here, the integers $H_p$ and $H_u$ represent the prediction and
control horizons, respectively. More specifically, $H_p$ defines the
number of future steps which have to be predicted, while $H_u$ is the
length of the future predicted control sequence. By definition, the
following relation holds: $H_u\leq H_p$.

In this way, the future evolution of the network can be predicted
$H_p$ steps ahead, as
\begin{equation}
  \label{eq: prediction system}
  X(k+1)=P_{X}x(k)+P_{U}U(k),
\end{equation}
with
\[\begin{array}{l}X^T(k+1)=
  \left[x^{T}(k+1),\ldots,x^{T}\left(k+H_p\right)\right]_{1\times H_p N},\\
  U^T(k)=\left[v^{T}(k),\ldots,v^{T}(k+H_u-1)\right]_{1\times H_u N},
\end{array}\]
$P_X^T=\left[P_\epsilon^{T},\ldots,\left(P_\epsilon^{H_p}\right)^{T}\right]_{H_p
  N \times N}$,
and the expression of $P_U$ given by (\ref{eq: PU}) in {\it Appendix \ref{app1}}.

Bearing in mind the goal of consensus protocol, i.e. eliminating the
disagreement of all the individuals of the network, we first calculate
the state difference of agents $i$ and $j$ in the network, $m~(1\leq
m\leq H_p)$ steps ahead, using the operator
\begin{equation}
  \label{eq: state-difference of a couple} \Delta
  x_{i,j}(k+m)\triangleq x_i(k+m)-x_j(k+m)=e_{i,j}x(k+m),
\end{equation}
where $e_{i,j}\triangleq e_i-e_j$ and $e_j\triangleq
[0,\cdots,0,\underbrace{1}_{j^{th}},0,\cdots,0]_{1\times N}$. Based on
(\ref{eq: state-difference of a couple}), the network state difference vector $m~(1\leq m\leq H_p)$
steps ahead can be defined by {\small
\[\begin{array}{l}\Delta x(k+m)\triangleq \left[ \Delta
    x^T_{1,2}(k+m),\ldots,\Delta x^T_{1,N}(k+m), \Delta
    x^T_{2,3}(k+m),\right. \\
  \left. \ldots,\Delta x^T_{2,N}(k+m),\ldots, \Delta
    x^T_{N-1,N}(k+m) \right] ^T_{N(N-1)/2\times 1}.
\end{array}\]
}

Consequently, the future evolution of the network's state difference
can be predicted $H_p$ steps ahead as follows:
\begin{equation}
  \label{eq: state difference iteration}
  \begin{array}{c}
    \Delta x(k+1)=ex(k+1),\\
    \vdots\\
    \Delta x(k+H_p)=ex(k+H_p)
  \end{array}
\end{equation}
with {\small $e\triangleq
  [e^T_{1,2},\ldots,e^T_{1,N},e^T_{2,3},\ldots,e^T_{2,N},\ldots,e^T_{N-1,N}]_{N(N-1)/2\times
    N}^T$}, $e_{i,j}\triangleq e_i-e_j$.

It then follows from (\ref{eq: state difference iteration}) that
{\small
  \begin{eqnarray}
    \Delta X(k+1) & \triangleq & \left[\Delta x(k+1)^T,\ldots, \Delta
      x(k+H_p)^T\right]^T_{H_pN(N-1)/2 \times 1} \nonumber\\
    & = & EX(k+1)=E(P_Xx(k)+P_UU(k)) \nonumber\\
    & = & P_{XE}x(k)+P_{UE}U(k)\label{eq:network_state_diff_dynamics}
\end{eqnarray}
} with $E\triangleq\mbox{diag}(e,\ldots,e)_{H_pN(N-1)/2 \times H_pN}$,
$P_{XE}\triangleq EP_X$ and $P_{UE}\triangleq EP_U$.

To solve the consensus problem, we first set the moving horizon
optimization index that defines the MPC consensus problem as follows:
\begin{equation}
  \label{eq: optimization index}
  J(k)=\left\| \Delta X(k+1)\right\|_Q^{2}+\left\| U(k)\right\|_R^{2},
\end{equation}
where $Q$ and $R$ are compatible real, symmetric, positive definite
weighting matrices, and $\|M\|_Q^2=M^TQM$. In general, the weighting
matrices can be set as
\begin{equation}
  \label{eq: QR} Q=qI_{H_pN(N-1)/2}~(q>0)~~  \mbox{and} ~~R=I_{H_uN}.
\end{equation}
In the optimization index (\ref{eq: optimization index}), the first
term penalizes the state difference between each pair of states over
the future $H_p$ steps, while the second term penalizes the additional
MPC control energy $v(k)$.  In order to minimize (\ref{eq:
  optimization index}), we compute $\partial J(k)/\partial U(k)=0$,
and obtain the optimal MPC action as:
\begin{equation}
  \label{eq: control law} v(k)=P_{MPC}x(k),
\end{equation}
where
\begin{eqnarray}
  P_{MPC}&=&-\left[I_N,\mathbf{0}_N,\ldots,\mathbf{0}_N\right]_{N\times H_u N}\nonumber \\
  &&\cdot(P_{UE}^TQP_{UE}+R)^{-1}P_{UE}^{T}QP_{XE},\label{eq: PMPC}
\end{eqnarray}
$I_N\in \mathbb{R}^{N\times N}$ and $\mathbf{0}_N\in
\mathbb{R}^{N\times N}$ are the identity and zero matrices of
dimension $N$, respectively.  The associated closed-loop dynamics can then
be written as 
\begin{equation}
  \label{eq: closed-loop system}
  \begin{array}{l}
    x(k+1)=(P_{\epsilon}+P_{MPC})x(k).
  \end{array}
\end{equation}

Interestingly, the proposed algorithm shows some consistency with the
routine protocol (\ref{eq: routine control}). More precisely, the
latter is solely based on the current state difference $\Delta
x_{i,j}(k)$ of each pair in the network while the former roots not
only in the current state difference $\Delta x_{i,j}(k)$ but also in
the future state difference $\Delta x_{i,j}(k+m)$, which constitutes
the main improvement of this method.


\subsection{Analysis}

For symmetric all-to-all link networks, it can be shown that $P_{MPC}$
and $P_\epsilon$ share the same eigenvectors. The following theorem
states this \textit{eigenvector conservation} property in more
details.

\begin{theorem}[Eigenvector conservation theorem]\label{Th:Eigenvalue
    conservation}
  Consider an $N$-node, all-to-all, symmetric network whose dynamics
  are described by (\ref{eq: closed-loop system with MPC law}) and
  (\ref{eq: control law}), and with the associated weighting matrices
  given by (\ref{eq: QR}). If $\lambda_{i}$ and $\eta_i$ denote the
  $i^{th}$ eigenvalue and the corresponding eigenvector of
  $P_{\epsilon}$, respectively, then $P_{MPC}$ (see (\ref{eq: PMPC}))
  satisfies the following relations:
  \begin{enumerate}
  \item $P_{MPC}\cdot\eta_i=\nu_{i}\cdot \eta_i$, where $\nu_i$ is the
    $i^{th}$ eigenvalue of $P_{MPC}$ corresponding to $\eta_{i}$;
  \item $P_{MPC}^T=P_{MPC}$.
  \end{enumerate}
\end{theorem}

\textit{Proof}:  See {\it Appendix \ref{app1}}.
$~~\hfill\blacksquare$

It follows from {\it Theorem~\ref{Th:Eigenvalue conservation}} that
\begin{equation}
  \label{eq: unchangable eigenvector of MPC}
  \lambda_i\left(P_\epsilon+P_{MPC}\right)=\lambda_i\left(P_\epsilon
  \right)+\lambda_i\left(P_{MPC}\right),~ i=1,\ldots,N,
\end{equation}
where $\lambda_i(A)$ denotes the $i^{th}$ eigenvalue of the matrix
$A$. For an arbitrary eigenvalue $\lambda_i~(1\leq i \leq N)$ of
$P_{\epsilon}$, the associated eigenvalue of $P_{MPC}$, $\nu_i$, is
given by
\begin{equation}
  \label{eq:MPC eigenvalue}
  \nu_i=-\gamma^*_1(H_u,H_p,\lambda_i,q),
\end{equation}
which can be calculated by (\ref{eq: Theorem1-3}) given in {\it
  Appendix~\ref{app1}}.

In particular, the eigenvalue of $P_{MPC}$ corresponding to the
trivial eigenvector $\textbf{\textit{1}}$ is $\nu_1=0$. This latter
property can be generalized to balanced networks as summarized in the
following corollary.

\begin{corollary}\label{trival eigenvector}
  Consider an all-to-all $N$-node network whose dynamics are described
  by (\ref{eq: closed-loop system with MPC law}) and (\ref{eq: control
    law}), and with associated weighting matrices given by (\ref{eq:
    QR}). If the network is balanced, then the matrix $P_{MPC}$ (see
  (\ref{eq: PMPC})) is balanced in the sense that
  $P_{MPC}\textit{\textbf{1}}=P_{MPC}^T\textit{\textbf{1}}=\textit{\textbf{0}}$
  with $\textit{\textbf{0}}\triangleq 0\cdot\textbf{\textit{1}}$.
\end{corollary}

\textit{Proof}: For balanced networks $P_{\epsilon}$ (see (\ref{eq:
    discrete-time network dynamics})), it is obvious that
  \begin{equation}
    \label{eq: trival eigenvector}
    P_{\epsilon}\textit{\textbf{1}}=P_{\epsilon}^{T}\textit{\textbf{1}}=\textit{\textbf{1}}.
\end{equation}
Using (\ref{eq: trival eigenvector}), a similar proof as the one given
to {\it Theorem~\ref{Th:Eigenvalue conservation}} shows that
$P_{XE}\textit{\textbf{1}}_{H_p N\times
  1}=EP_X\textit{\textbf{1}}_{H_p N\times
  1}=E[\textit{\textbf{1}},\ldots,\textit{\textbf{1}}]^T_{H_pN\times1}=[0,\ldots,0]^T_{H_pN(N+1)/2\times
  1}$, which leads to
$P_{MPC}\textit{\textbf{1}}=\textit{\textbf{0}}$. In other words, the
eigenvalue of $P_{MPC}$ associated with the trivial eigenvector
\textit{\textbf{1}} is $0$.  Moreover, it is easy to see that {\small
\begin{equation}\label{eq: corollary 1-2}
  \textit{\textbf{1}}^T\left[I_N,\mathbf{0}_N,\ldots,\mathbf{0}_N\right]_{N\times
    H_u N}=\left[\textit{\textbf{1}}^T,\textit{\textbf{0}}^T,\ldots,\textit{\textbf{0}}^T\right](P_{UE}^TQP_{UE}+R).
\end{equation}
}

Since $P_{UE}^TQP_{UE}+R$ is invertible, it follows from (\ref{eq:
  corollary 1-2}) that
\begin{equation}\label{eq: corollary-3}
  \begin{array}{c}\textit{\textbf{1}}^T\left[I_N,\mathbf{0}_N,\ldots,\mathbf{0}_N\right]_{N\times
      H_u N}(P_{UE}^TQP_{UE}+R)^{-1}\\
    =\left[\textit{\textbf{1}}^T,\textit{\textbf{0}}^T,\ldots,\textit{\textbf{0}}^T\right].
  \end{array}
\end{equation}
Substituting (\ref{eq: corollary-3}) into (\ref{eq: PMPC}) yields that
$\textit{\textbf{1}}^TP_{MPC}=\textit{\textbf{0}}^T$. Thus, $P_{MPC}$
is balanced in the sense that
$\textit{\textbf{1}}^TP_{MPC}=\textit{\textbf{0}}^T$ and
$P_{MPC}\textit{\textbf{1}}=\textit{\textbf{0}}.$
$~~\hfill\blacksquare$

A direct consequence of \textit{Corollary~\ref{trival eigenvector}} is
that the state matrix of the MPC protocol $P_{\epsilon}+P_{MPC}$ is
balanced in the sense that
$(P_{\epsilon}+P_{MPC})^T\textit{\textbf{1}}=(P_{\epsilon}+P_{MPC})\textit{\textbf{1}}=\textit{\textbf{1}}$.

Note that balanced networks are more general than symmetric networks,
thus {\it Corollary~\ref{trival eigenvector}} is more general than
{\it Theorem~\ref{Th:Eigenvalue conservation}}.

Based on {\it Theorem~\ref{Th:Eigenvalue conservation}} and {\it
  Corollary~\ref{trival eigenvector}}, we give hereafter necessary and
sufficient conditions guaranteeing asymptotic convergence to the
average-consensus for the proposed MPC protocol (see (\ref{eq:
  closed-loop system with MPC law}) and (\ref{eq: control law})).

\begin{lemma}\label{Th: partial link lemma}
  For any matrix $W\in \mathbb{R}^{N\times N}$, the equation
  \begin{equation}
    \label{eq: lemma3-1}
    \lim_{k\rightarrow
      \infty}~W^k=\textbf{\textit{1}}\textbf{\textit{1}}^T/N
\end{equation}
holds if and only if either assumptions \textit{A1} and \textit{A2}
hold or assumptions \textit{A1} and \textit{A3} hold

\begin{enumerate}
\item[\textbf{\textit{A1}}:]
  \begin{equation}
    \label{eq: lemma3-2}
    \textbf{\textit{1}}^TW=\textbf{\textit{1}}^T~~~ \mbox{and}~~~
    W\textbf{\textit{1}}=\textbf{\textit{1}};
\end{equation}

\item[\textbf{\textit{A2}}:]
  \begin{equation}
    \label{eq: lemma3-3}
    \rho(W-\textbf{\textit{1}}\textbf{\textit{1}}^T/N)<1,
\end{equation}
where $\rho(\cdot)$ denotes the spectral radius of a matrix;

\item[\textbf{\textit{A3}}:] the matrix $W$ has a simple eigenvalue at
  $1$ and all its other eigenvalues in the open unit circle.
\end{enumerate}
\end{lemma}

\textit{Proof}:  
  The property ``Equation (\ref{eq: lemma3-1}) $\Leftrightarrow$
  \textit{Assumptions} \textit{A1} and \textit{A2} hold'' has been
  proven in \cite{xi04}. The proof of the the property ``Equation
  (\ref{eq: lemma3-1}) $\Leftrightarrow$ \textit{A1} and \textit{A3}
  hold'' is provided in \textit{Appendix \ref{app2}}.
$~~\hfill\blacksquare$

Based on \textit{Lemma~\ref{Th: partial link lemma}} and
\textit{Corollary~\ref{trival eigenvector}}, some necessary and
sufficient condition of the proposed MPC protocol (see (\ref{eq:
  closed-loop system with MPC law}) and (\ref{eq: control law})) is
provided as follows.

\begin{theorem}\label{Th: equilibrium} For the closed-loop system
  (\ref{eq: closed-loop system}) associated with an $N$-node
  all-to-all, balanced network whose dynamics are described by
  (\ref{eq: closed-loop system with MPC law}) and (\ref{eq: control
    law}), the system state $x(k)$ asymptotically converges to the
  equilibrium point $\bar{x}(0)\textit{\textbf{1}}$ with
  $\bar{x}(0)\triangleq 1/N\sum_{i=1}^{N}x_i(0)$ if and only if either
  of the following two assumptions holds

  \begin{enumerate}
  \item[\textit{\textbf{A4}}:]
    $\rho(P_{\epsilon}+P_{MPC}-\textbf{\textit{1}}\textbf{\textit{1}}^T/N)<1$;

  \item[\textit{\textbf{A5}}:] the matrix $P_{\epsilon}+P_{MPC}$ has a
    simple eigenvalue at $1$ and all its other eigenvalues in the open
    unit circle.
  \end{enumerate}
\end{theorem}

\textit{Proof}:  The asymptotic state value $x^*$ of the closed-loop system (\ref{eq:
    closed-loop system}) is given by
  \begin{equation}
    \label{eq:app2-0}
    x^*=\lim_{k\rightarrow \infty}(P_{\epsilon}+P_{MPC})^kx(0).
  \end{equation}
  Let $W=P_{\epsilon}+P_{MPC}$. Then, \textit{Corollary \ref{trival
      eigenvector}} implies that (\ref{eq: lemma3-2}) holds. It then
  follows from {\it Lemma~{\ref{Th: partial link lemma}}} that

  \[x^*=\lim_{k\rightarrow
    \infty}(P_{\epsilon}+P_{MPC})^kx(0)=\textbf{\textit{1}}\textbf{\textit{1}}^T/N\cdot
  x(0)=\bar{x}(0)\textit{\textbf{1}}\] if and only if either of the
  two \textit{Assumptions} \textit{A4} or \textit{A5} holds.
$~~\hfill\blacksquare$

Bearing in mind the balanced feature of $P_{MPC}$, it can be proven,
using the \emph{Ger$\check{s}$gorin Disc Theorem} \cite{ho90}, that
(i) for fixed values of $\epsilon$, the MPC protocol can compress the
corresponding cluster of eigenvalues and drive it to approach the
origin; (ii) the MPC protocol can significantly expand the feasible
range of $\epsilon$. Details of these statements are given in the
following lemma and theorem.

\begin{lemma}[Ger$\check{s}$gorin Theorem
  \cite{ho90}]\label{Th:Gersgorin Theorem}
  Let $A=\left\{a_{ij}\right\}_{i,j=1,\ldots,N}\in \mathbb{R}^{N\times
    N}$, and let $R_i^{'}(A)\triangleq \sum_{j=1,j\neq i}^N |a_{ij}|,~
  i=1,2,\ldots,N$. Then all the eigenvalues of $A$ are located in the
  union of the $N$ discs
  \begin{equation}
    \label{Gersgorin Theorem} \cup_{i=1}^n \left\{z \in \mathbb{C}:
      |z-a_{ii}|\leq R_i^{'}(A)\right\}.
  \end{equation}
\end{lemma}

\begin{figure}[htb]
  \centering \leavevmode
  {\includegraphics[width=5.2cm]{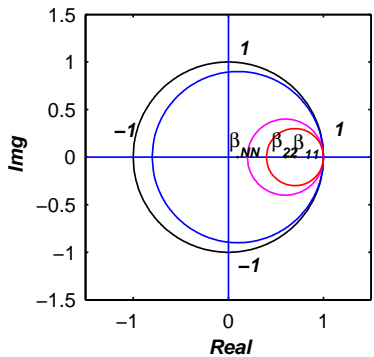}}
  \centering \caption{(Color online) Illustration of {\it
      Theorem~\ref{Th: eigenvector cluster improvement}}}
  \label{fig: theorem 3}
\end{figure}

\begin{theorem}\label{Th: eigenvector cluster improvement}

  Consider an all-to-all, strongly connected, balanced network whose
  dynamics are described by (\ref{eq: closed-loop system with MPC
    law}) and (\ref{eq: control law}). Denote each entry of
  $P_{\epsilon}+P_{MPC}$ by $\beta_{ij},~i,j=1,\ldots,N.$ Under the
  assumptions that $\beta_{ij}\geq 0~(i\neq j)$ and $\beta_{ii} \in
  (0,1]$, the network state $x(k)$ will asymptotically converge to
  $\bar{x}(0)\textit{\textbf{1}}$. In other words, average-consensus
  will be achieved asymptotically.
\end{theorem}

\textit{Proof}:  It follows from {\it Corollary~\ref{trival eigenvector}} that
  $(P_{\epsilon}+P_{MPC})\textit{\textbf{1}}=\textit{\textbf{1}}$. Moreover,
  since $\beta_{ij}\geq 0~(i\neq j)$, one has
  $R_i^{'}(P_{\epsilon}+P_{MPC})=\sum_{j=1,j\neq i}^N
  \beta_{ij}=1-\beta_{ii}$. From {\it Lemma~\ref{Th:Gersgorin
      Theorem}}, one obtains that all the eigenvalue of
  $P_{\epsilon}+P_{MPC}$ are located in the union of the $N$ discs
  \begin{equation}
    \label{eq: Theorem 3-2} \cup_{i=1}^n \left\{z \in \mathbb{C}:
      |z-\beta_{ii}|\leq 1-\beta_{ii}\right\},
  \end{equation}
  as shown in Fig.~\ref{fig: theorem 3}.  Therefore, if $\beta_{ii}
  \in (0,1]$, then all the eigenvalues of the state matrix
  $P_{\epsilon}+P_{MPC}$ are in the open unit circle, except the
  simple eigenvalue $1$. Moreover, if the $\beta_{ii}$, $i=1,\ldots,N$
  are arranged in non-ascending order,
  i.e. $0<\beta_{NN}\leq\beta_{N-1,N-1}\leq \ldots\leq \beta_{11}\leq
  1$, as shown in Fig.~\ref{fig: theorem 3}, then the union of the $N$
  discs (\ref{eq: Theorem 3-2}) equals
  \begin{equation}
    \label{eq: Theorem 3-3}
    \left\{z \in \mathbb{C}:
      |z-\beta_{NN}|\leq 1-\beta_{NN}\right\}.
  \end{equation}
  Thus, taking {\it Theorem~\ref{Th: equilibrium}} into consideration,
  one can conclude that average-consensus will be achieved asymptotically.
$~~\hfill\blacksquare$

\begin{remark}
  The assumptions $\beta_{ij}\geq 0~(i\neq j)$ and $\beta_{ii} \in
  (0,1]$ can be easily satisfied in general.  Indeed, it can be numerically
  checked that in the $3$-dimensional space spanned by the parameters
  $H_u$, $H_p$, and $q$, there is a fairly large region in which these
  two assumptions are satisfied (such as the region corresponding to
  the common parameter settings $H_u\in [1,10]$, $H_p\in [H_u,10]$,
  and $q\in [0.1,10]$). Moreover, the circle determined by (\ref{eq:
    Theorem 3-3}) shrinks with increasing values of
  $\beta_{NN}~(\beta_{NN} \in (0,1))$, and the conditions given in
  {\it Theorem~{\ref{Th: eigenvector cluster improvement}}} thus
  become more and more restrictive. Fortunately, for the common
  parameter settings $H_u\in [1,10]$, $H_p\in [H_u,10]$, and $q\in
  [0.1,10]$, it can be shown by simulations that $\beta_{NN}$ is
  generally very close to zero when $\epsilon \leq 20/d_{\max}$,
  thereby reducing the conservativeness of the conditions of {\it
    Theorem~{\ref{Th: eigenvector cluster improvement}}}. In this way,
  the feasible range of the sampling period $\epsilon$ is
  significantly broadened. Indeed, when the
  routine state matrix $P_{\epsilon}$ is not discrete-time Hurwitz
  (which happens for $1/d_{\max} \leq \epsilon$ \cite{sa07}), the MPC
  state matrix $P_{\epsilon}+P_{MPC}$ will still be discrete-time
  Hurwitz if $1/d_{\max}\leq \epsilon \leq \bar{\epsilon}$ where
  $\bar{\epsilon}$ is a certain threshold, typically much larger than
  $1/d_{\max}$. As shown in {\it Theorem~\ref{Th: eigenvector cluster
      improvement}}, this is due to the fact that the MPC consensus
  protocol is able to compress the cluster of eigenvalues and drive it
  to approach the origin when $\epsilon$ grows larger than the
  threshold value $1/d_{\max}$, as will be illustrated later, in the
  case study.
\end{remark}

\subsection{Case study}
To vividly illustrate the advantages of the MPC consensus protocol, we
present some simulation results comparing the convergence speeds
obtained using the routine protocol given in (\ref{eq: routine
  control}) and the proposed MPC protocol given in (\ref{eq:
  closed-loop system with MPC law}), (\ref{eq: control law}) for the
particular case of all-to-all, balanced networks.


We first consider an all-to-all, symmetric network of $10$ nodes.
Since the objective is to reach average-consensus, the instantaneous
disagreement index is typically set as $D(k)\triangleq
\|x(k)-\textbf{\textit{1}}\bar{x}(0)\|_2^{2} $ and the
\textit{consensus steps} $T_c$ can be defined as the running steps
required for $D(k)$ to reach a specified neighborhood of the origin
and stay therein afterwards, i.e.
\begin{equation}
  \label{eq: convergent step definition} T_c(D_c)=\min \left\{T \in
    \mathbb{R}^{+} : D(k)\leq D_c,\, \forall k\geq T\right\},
\end{equation}
where $D_c$ is a positive number defined as the \textit{consensus
  threshold}, and $\|x\|_2=(x^Tx)^{1/2}$. It is clear that
$1/T_c(D_c)$ gives a reasonable measurement of the \textit{consensus
  speed}.

\begin{figure}[htp]
\centering \leavevmode
\begin{tabular}{cc}
\hspace*{-0.3cm}
\resizebox{4.2cm}{!}{\includegraphics[width=11.2cm]{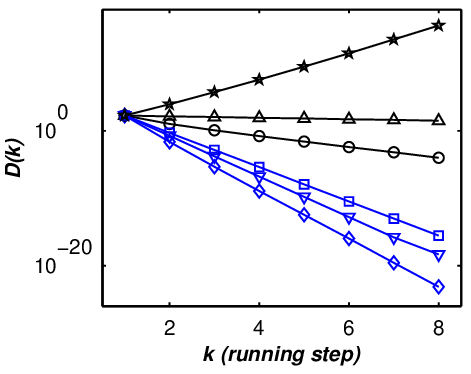}}
&
\resizebox{4.2cm}{!}{\includegraphics[width=11.2cm]{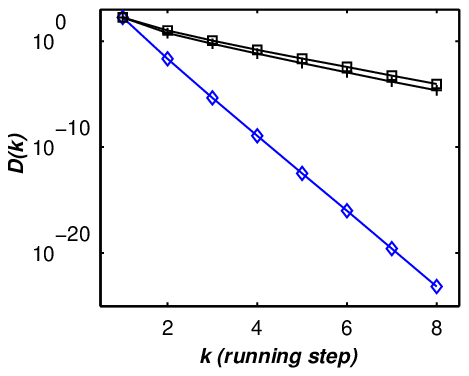}}\\
{\scriptsize (a) } &  {\scriptsize (b) }
\end{tabular}
\caption{(Color online) \textbf{(a)} Comparison of the routine (black
  curves) and MPC (blue curves) protocols on the time evolution of the
  instantaneous disagreement index $D(k)$ corresponding to an
  all-to-all, symmetric network with different sampling periods
  $\epsilon$. Here, $l_{ij}=-1~(i\neq j)$ and $l_{ii}=N-1$. Case~1:
  $\epsilon=1/9=1/d_{\max}$, $\lozenge$: MPC, {\scriptsize$\bigcirc$}:
  routine protocol; Case~2: $\epsilon=0.03<1/d_{\max}$,
  $\bigtriangledown$: MPC, $\bigtriangleup$: routine protocol; Case~3:
  $\epsilon=2.00\gg 1/d_{\max}$, $\square$: MPC, $\bigstar$: routine
  protocol.  ~~\textbf{(b)} Comparison of FDLA (black curves) and MPC
  (blue curves) protocols on a symmetric network with each
  off-diagonal entry $l_{ij}=l_{ji}~(i\neq j)$ selected randomly in
  $[-1,0]$. Here $\lozenge$: MPC; {$+$}: FDLA1; $\square$: FDLA2,
  $\epsilon=0.20$ and $d_{\max}\simeq 5$. In both (a) and (b), the MPC
  parameters are $N=10$, $H_u=2$, $H_p=4$, $q=2$, and the initial
  states $x_i(0),~i=1,\ldots,N$ are selected randomly in $[0,15]$. }
\label{fig:convergence comparison}
\end{figure}

As shown in Fig.~\ref{fig:convergence comparison}(a), the addition of
the predictive mechanism defined in (\ref{eq: control law}), yields a
drastic increase in convergence speed towards consensus. In
particular, for $\epsilon \ll 1/d_{\max}$ ($d_{\max}=
\max_{i}\left(l_{ii}\right)$), the convergence speed is increased more
than $20$ times by using the proposed MPC protocol.  Furthermore, even
when the routine convergence conditions are violated, i.e. $\epsilon >
1/d_{\max}$, it is observed that the MPC consensus protocol still
allows asymptotic convergence with a high-speed.

To further illustrate the advantages of the MPC protocol, comparison
results of this latter protocol with the {\it fastest distributed
  linear averaging} (FDLA) protocols proposed in \cite{xi04} are
presented in Fig.~\ref{fig:convergence comparison}(b). More precisely,
there are mainly three kinds of FDLA: if the dynamics of the
considered network is determined by $x(k+1)=Wx(k)$, then
\begin{itemize}
\item in FDLA1, namely the \textit{best constant} method, $W=I_N-\alpha
  L$ with $\alpha=2/(\lambda_2(L)+\lambda_N(L))$. FDLA1 is applicable only to symmetric networks;
\item in FDLA2, namely the \textit{maximum-degree weight} method,
  $W=I_N-\alpha L$ with $\alpha=1/d_{\max}$. FDLA2 is applicable only to symmetric networks;
\item in FDLA3, namely the \textit{spectral norm minimization} method,
  $W$ is the solution of the following spectral norm minimization
  problem:
\begin{equation}
\label{eq: FDLA3}
\begin{array}{c}
  \min_W~~~\Vert W-\textbf{\textit{1}}\textbf{\textit{1}}^T/N \Vert\\
%
s.t.~~~W \in \mathcal{E},~
~\textbf{\textit{1}}^TW=\textbf{\textit{1}}^T, ~W
\textbf{\textit{1}}=\textbf{\textit{1}}.
\end{array}
\end{equation}
\end{itemize}
It has been proven that the optimization problem (\ref{eq: FDLA3}) in
FDLA3 is convex, thus FDLA3 can yield a global optimum $W^*$ for the
consensus problem\cite{xi04}. Additionally, we note that FDLA2 is
typically slower than FDLA1. The convergence properties of both FDLA1
and FDLA2 are described in \cite{xi04}.

It can be observed from Fig.~\ref{fig:convergence comparison}(b) that
the consensus speed of the proposed MPC protocol is much faster than
that of FDLA1 and FDLA2, which is due to the larger degree of freedom
allowed by the state matrix $W=P_{\epsilon}+P_{MPC}$. As to FDLA3, one
can get that the optimum $W^*=
\textbf{\textit{1}}^T\textbf{\textit{1}}/N$ for all-to-all link
networks. As a consequence, $x(1)=Wx(0)=\textbf{\textit{1}}\bar{x}$,
namely, the disagreement index $D(k)$ will be zero in just one step,
making FDLA3 the fastest possible consensus algorithm for all-to-all link
networks. We have proven in {\it Theorem~\ref{Th: equilibrium}} that
the state matrix of the MPC protocol satisfies $\lim_{k\rightarrow
  \infty}(P_{\epsilon}+P_{MPC})^k=W^*$, provided that the matrix
$P_{\epsilon}+P_{MPC}$ has a simple eigenvalue at $1$ and all its
other eigenvalues in the open unit circle.  Furthermore, considering
all-to-all link networks, the constraints imposed in (\ref{eq: FDLA3})
are all fulfilled for $W=P_{\epsilon}+P_{MPC}$. Interestingly, we
observe in Fig.~\ref{fig:convergence comparison} that
$(P_{\epsilon}+P_{MPC})^k$ quickly approaches
$W^*=\textbf{\textit{1}}^T\textbf{\textit{1}}/N$, which nicely
illustrates the results presented in {\it Theorem~\ref{Th:
    equilibrium}}.

To further compare the convergence performances of the MPC and the
routine protocols, we now generalize the above all-to-all, symmetric
networks to an all-to-all, asymmetric, balanced networks with
off-diagonal entries $a_{ij},~i\neq j$ selected randomly from
$[-1,0]$. The corresponding time evolution of the instantaneous
disagreement index $D(k)$ is fairly similar to the one presented in
Fig.~\ref{fig:convergence comparison} for a symmetric all-to-all
topology and is thus omitted here. Nevertheless, we show their
associated \textit{ultrafast-convergence} probabilities $P_c(D_c)$
with respect to $\epsilon$ in Fig.~\ref{fig:all-to-all-link converge
  probability}(a). Here, $P_c(D_c)$ denotes the ratio of ultrafast-convergence runs over $500$ runs for each $\epsilon$. A convergence
run is considered ultrafast if the consensus threshold $D_c=0.01$ (see
(\ref{eq: convergent step definition})) can be reached within $100$
steps. It can be observed that, with the assistance of a predictive
mechanism, the ultrafast-convergence probability is significantly
increased for each fixed value of $\epsilon$.  Furthermore, the
maximum feasible ultrafast-convergence sampling period $\epsilon$ is also sharply
increased (more than $40$ times from our simulation results).

To further study the effects of the predictive mechanism and analyze
its impact on the feasible convergence range of sampling periods
$\epsilon$, we examine in Fig.~\ref{fig:all-to-all-link converge
  probability}(b) the evolution of the consensus steps $T_c(0.01)$ of
these two strategies with respect to increasing values of
$\epsilon$. In this comparison, for each value of $\epsilon$, $T_c$
denotes the average of the consensus steps corresponding to the
successful convergence runs over a total of $500$ runs. It can be
seen that, compared with the routine protocol, the MPC protocol
allows for a significant increase in the consensus speed
$1/T_c(0.01)$ (by a factor between $6$ and $20$ in our simulation
results).

\begin{figure}[htp]
\centering \leavevmode
\begin{tabular}{cc}
\hspace*{-0.3cm}
\resizebox{4.2cm}{!}{\includegraphics[width=11.2cm]{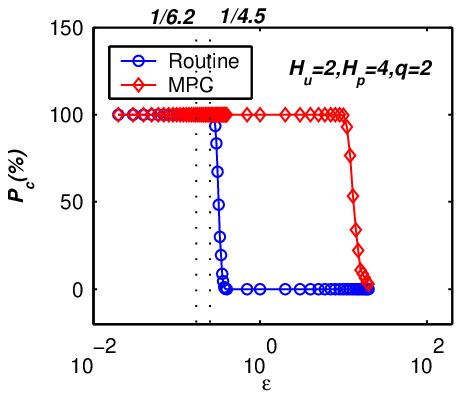}}
&
\resizebox{4.25cm}{!}{\includegraphics[width=11.2cm]{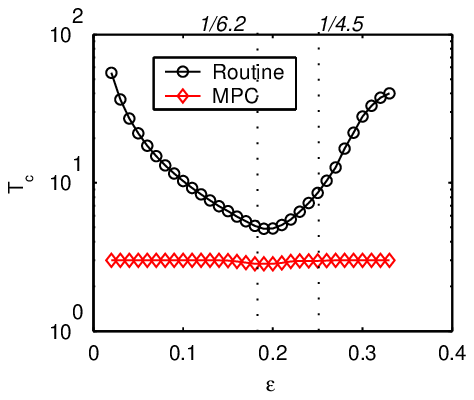}}\\
{\scriptsize (a)} &  {\scriptsize (b)}
\end{tabular}
\caption{(Color online) (a): All-to-all link network's ultrafast-convergence probability ($P_c(0.01)$); (b): consensus steps
  ($T_{c}(0.01)$). Comparison is addressed between the MPC (red
  curves) and routine (blue curves) protocols with different sampling
  periods $\epsilon$ and $500$ independent runs for each value of
  $\epsilon$. In these simulations, $H_u=2$, $H_p=4$, $q=2$, the
  consensus threshold $D_c=0.01$, each entry $l_{ij}(i\neq j)$ of $L$
  is selected randomly in $[-1,0]$, and $x_i(0)~(i=1,\ldots,N)$ is
  selected randomly in $[0,15]$. The corresponding values of
  $d_{\max}$ lie in the range $[4.5,6.2]$. The vertical dotted lines
  correspond to the minimum and maximum values of
  $1/d_{\max}$.} \label{fig:all-to-all-link converge probability}
\end{figure}

\begin{figure}[htp]
\centering \leavevmode
\begin{tabular}{cc}
\resizebox{4.3cm}{!}{\includegraphics[width=11.2cm]{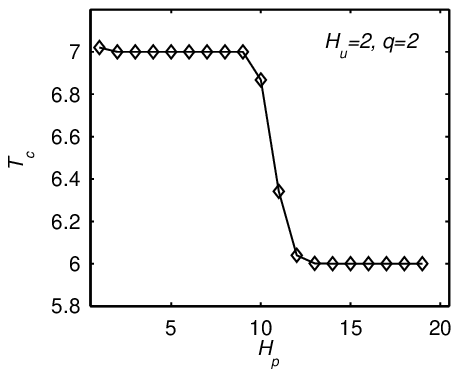}}
&
\resizebox{4.35cm}{!}{\includegraphics[width=11.2cm]{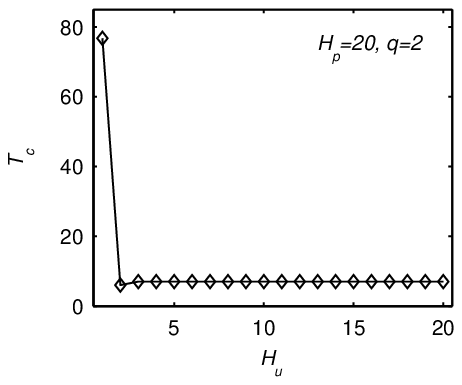}}
 \\
{\scriptsize (a)} &  {\scriptsize (b)} \\
\resizebox{4.3cm}{!}{\includegraphics[width=11.2cm]{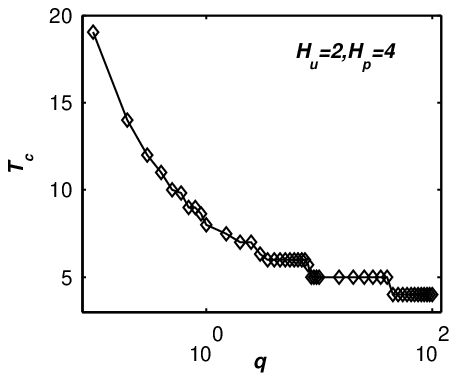}}
&
\resizebox{4.35cm}{!}{\includegraphics[width=11.2cm]{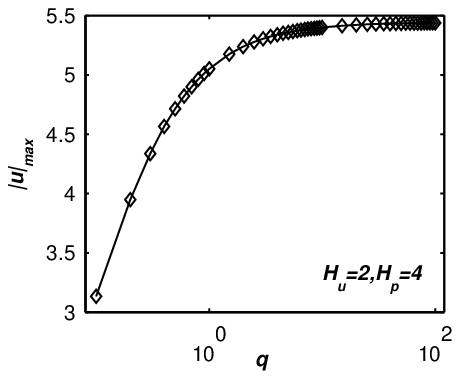}}\\
{\scriptsize (c)} &  {\scriptsize (d)}
\end{tabular}
\caption{(Color online) Evolution of the consensus steps
  ($T_c(10^{-7})$) of the MPC consensus protocol with respect to the
  control horizon $H_u$, the prediction horizon $H_p$ and the
  weighting parameter $q$. Here, the threshold $D_c=10^{-7}$, and the
  sampling period $\epsilon=0.05$.}
\label{fig:consensus steps of MPC}
\end{figure}

Since there are three crucial parameters determining the performances
of the proposed MPC protocol, namely $H_p$, $H_u$ and $q$ (see
(\ref{eq: QR})), statistical simulations have been carried out to
investigate their individual influence on the convergence speed. The
results of these simulations are shown in Fig. \ref{fig:consensus
  steps of MPC}. One can observe that (i) the consensus speed
$1/T_c(10^{-7})$ is enhanced with increasing values of the parameters
$H_p$ (see Fig.~\ref{fig:consensus steps of MPC}(a)) or $q$ (see
Fig.~\ref{fig:consensus steps of MPC}(c)); (ii) a global maximum of
$1/T_c(10^{-7})$ exists at a value of $H_u$ (see
Fig.~\ref{fig:consensus steps of MPC}(b)). Furthermore, $T_c(10^{-7})$
remains stable when $H_u$ exceeds a specified threshold (see also
Fig.~\ref{fig:consensus steps of MPC}(b)).

Generally speaking, an increase in the values of $H_p$, $H_u$ and $q$,
can improve the overall consensus performance. However, when their
corresponding values exceed some thresholds, this improvement becomes
negligible. Meanwhile the control efforts and computational burdens
are still drastically increased (see Fig.~\ref{fig:consensus steps of
  MPC}(d)). Consequently, taking into consideration both computational
complexity and consensus speed, one can find optimal parameter values
according to the statistical simulation results depicted in
Fig.~\ref{fig:consensus steps of MPC}.


\begin{figure}[htp]
\centering \leavevmode
\begin{tabular}{cc}
\resizebox{4.3cm}{!}{\includegraphics[width=11.2cm]{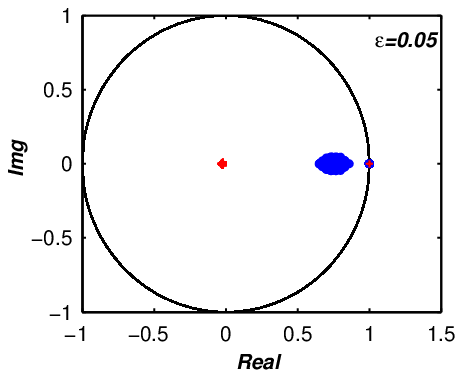}}
&
\resizebox{4.25cm}{!}{\includegraphics[width=11.2cm]{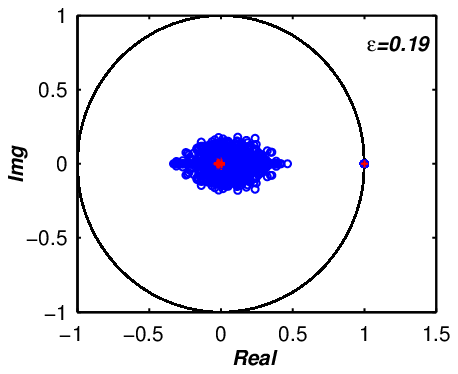}}
 \\
{\scriptsize (a)} &  {\scriptsize (b)} \\
\resizebox{4.3cm}{!}{\includegraphics[width=11.2cm]{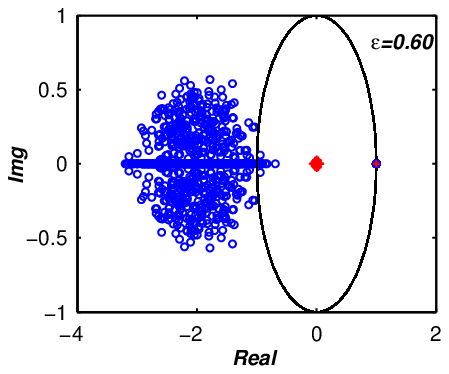}}
&
\resizebox{4.3cm}{!}{\includegraphics[width=11.2cm]{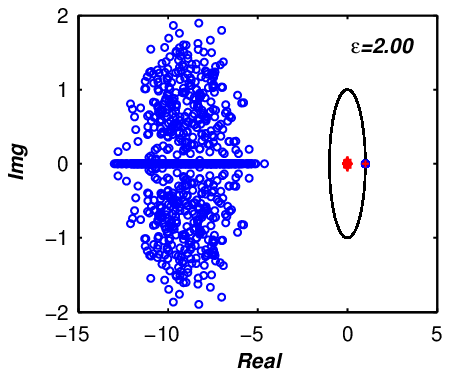}}\\
{\scriptsize (c)} &  {\scriptsize (d)}
\end{tabular}
\caption{(Color online) Eigenvalue distributions with respect to the
  sampling period $\epsilon.$ Blue {\scriptsize$\bigcirc$} and red $+$
  denote the eigenvalue distributions of $P_{\epsilon}$ and
  $P_{\epsilon}+P_{MPC}$ over $100$ runs, respectively. The black
  circle represents the unit circle in the complex plane. Here,
  $H_u=2$, $H_p=4$, $q=2$, and the entries $l_{ij},~j\neq i$ of $L$
  are chosen randomly from $[-1,0]$. The corresponding values of
  $d_{\max}$ lie in the interval $[4.5,6.2]$ in this $100$-run
  simulation.}
\label{fig:eigenvalue distribution of MPC}
\end{figure}

It is well known that the eigenvalue distribution of the state matrix
associated with the considered consensus protocol,
i.e. $P_{\epsilon}+P_{MPC}$, is closely linked to its
performances. Therefore, to further study the proposed MPC protocol
and better understand its ultrafast-convergence capability, we display
the eigenvalue distributions of $P_{\epsilon}$ and
$P_{\epsilon}+P_{MPC}$ in Fig.~\ref{fig:eigenvalue distribution of
  MPC}. These statistical simulations have been realized by
considering $100$ different runs for a balanced, asymmetric network
with entries $l_{ij},~i\neq j$ of the Laplacian matrix $L$ (appearing
in the definition of the matrix $P_{\epsilon}$, see (\ref{eq:
  discrete-time network dynamics})) randomly selected from
$[-1,0]$. For each run, we have considered $4$ different sampling
period values for $\epsilon$. Based on these simulation results, an
interesting phenomenon can be observed: the eigenvalue cluster of the
matrix $P_{\epsilon}+P_{MPC}$ is always more compact around the origin
than the counterpart cluster of $P_{\epsilon}$.

With these eigenvalue distributions, we can visualize the advantages
of the MPC protocol more lively. Indeed, since the eigenvalue cluster
of $P_{\epsilon}+P_{MPC}$ is always much smaller and closer to the
origin of the complex plane than that of $P_{\epsilon}$, the MPC
protocol generally has better consensus performance. When $\epsilon <
1/d_{\max}$ ($\epsilon=0.05$, see Fig.~\ref{fig:eigenvalue
  distribution of MPC}(a)), the two eigenvalue clusters both remain
inside the asymptotic stability region, i.e. the unit circle in the
complex plane. However, since each eigenvalue of
$P_{\epsilon}+P_{MPC}$ is much closer to the origin, the consensus
speed is sharply increased in the case of the MPC
protocol. Furthermore, when $\epsilon > 1/d_{\max}$ ($\epsilon=0.60,
2.00$, see Figs.~\ref{fig:eigenvalue distribution of MPC}(c) and (d)),
some of the eigenvalues of $P_{\epsilon}$ escape the unit circle,
making the disagreement function diverge, whereas all the eigenvalues
of $P_{\epsilon}+P_{MPC}$ remain inside the unit circle, guaranteeing
its asymptotic convergence.  Finally, still worth mentioning is that
the trivial eigenvalue of $P_{\epsilon}+P_{MPC}$, which corresponds to
the eigenvector $\textit{\textbf{1}}$, always remains at $1$
irrespective of the value of $\epsilon$.


In summary, the advantages of the proposed MPC protocol are twofold:
(i) enhancing the consensus speed associated with a fixed value of the
sampling period $\epsilon$; (ii) enlarging the range of feasible
convergence sampling periods. From the natural science point of view,
the proposed MPC protocol, and especially its predictive mechanism,
can be used to explain why individuals of bio-groups do not
communicate with each others very frequently but only occasionally in
a suitable manner during the whole dynamic process. From the
industrial application point of view, due to the enlargement of the
feasible range of the sampling periods, the use of the proposed MPC
protocol allows for a significant reduction of the communication costs
required to achieve a desired consensus speed.

\section{Model predictive consensus protocol for partial link networks}
\label{Model predictive consensus protocol for partial link
networks}

In this section, we further compare the routine and the MPC consensus
protocols in the more general case of \textit{partial link} networks
in which a pair of nodes is not necessarily connected. If we allow the
topology of the network to be changed, in other words, if we allow the
addition of some new edges in the graph, then the MPC protocol
({\ref{eq: control law}}) proposed for all-to-all link networks can be
used, and thus {\it Theorems~\ref{Th:Eigenvalue
    conservation}--\ref{Th: eigenvector cluster improvement}} and {\it
  Corollary~\ref{trival eigenvector}} remain valid. However, if we
assume that the initial sparsity structure of the network is fixed,
i.e. no new edge can be added in the graph, then we need to revise the
MPC protocol (\ref{eq: control law}) to obtain a new one suitable for
this type of scenario. In the following sections, we first introduce a
revised partial link, sparsity-preserving MPC consensus
protocol. Afterwards, to support this latter protocol, we derive some
necessary and sufficient condition guaranteeing its asymptotic
convergence towards average-consensus. Finally, we provide simulation
results for balanced partial link networks that show the advantages of
this MPC protocol.

\subsection{Algorithm}


In partial link networks, each node is allowed to communicate solely
with its neighbors. To ensure that the communication structure of
the network $G=(\mathcal{V},\mathcal{E},A)$ is preserved, one can
slightly revise the additional MPC term $v(k)$ appearing in
(\ref{eq:
  closed-loop system with MPC law}) to $v(k)=\Theta x(k)$, which leads
to the following linear network dynamics
\begin{equation}
  \label{eq: closed-loop system with MPC law and partial link}
  x(k+1)=P_{\epsilon}x(k)+\Theta x(k),
\end{equation}
where $\Theta= \left\{\theta_{ij}\right\}_{i,j=1,\ldots,N}\in
\mathbb{R}^{N \times N}$ satisfies the following relations:
\begin{enumerate}
\item the matrix $\Theta$ has the same sparsity structure as that of
  the network, i.e.
  \begin{equation}
    \label{eq: not change the topology}
    (i,j)\notin\mathcal{E},~i\neq j\Rightarrow \theta_{ij}=0;
  \end{equation}

\item the matrix $\Theta$ will not change the symmetric property of
  the network, i.e.
  \begin{equation}\label{eq: not change the symmetry}
    A^T=A\Rightarrow\Theta^T=\Theta;
  \end{equation}
\item $\Theta$ is balanced in the sense that
  \begin{equation}
    \label{eq: condition of Theta}
    \Theta\textit{\textbf{1}}=\Theta^T\textit{\textbf{1}}=\textbf{\textit{0}}.
  \end{equation}
\end{enumerate}
Note that these constraints are given to ensure feasibility of
consensus by the proposed MPC protocol, which will be proven later.
Fortunately, with these conditions, the degrees of freedom of $\Theta$
are reduced and the computational complexity of the proposed MPC
algorithm is thus reduced.

By iterating the dynamic model given in (\ref{eq: closed-loop system
  with MPC law and partial link}), the evolution of the state of the
$j^{th}$ agent can be predicted as
\begin{equation}
  \label{eq: partial link prediction}
  \begin{array}{l}
    x_j(k+m)=e_j(P_\epsilon+\Theta)^{m}x(k),
    \\~~~~~~~~~~~~~~~m=1,\cdots,H_p; ~j=1,\cdots,N,
  \end{array}
\end{equation}
with $e_j\triangleq
[0,\cdots,0,\underbrace{1}_{j^{th}},0,\cdots,0]_{1\times N}$.  Similar
to the all-to-all link case, the $m$-steps-ahead state difference of
the $i^{th}$ and $j^{th}$ agents can be calculated by (\ref{eq:
  state-difference of a couple}). However, in this partial link case,
the $m$-steps-ahead network state difference vector, i.e.  $\Delta
x(k+m)$, is solely composed of the state differences of the
neighboring pairs, i.e. $\Delta x(k+m)=\left(\mbox{col}\left(\Delta
    x_{i,j}(k+m)|(i,j)\in \mathcal{E}\right)\right)^T$. Note that if
both $(i,j)$ and $(j,i)$ are in $\mathcal{E}$, only $x_{i,j}(k+m),
~i>j$ will appear in $\Delta x(k+m)$.


Accordingly, the future evolution of the state difference can be
predicted $H_p$ steps ahead:
\begin{equation}
  \label{eq: state difference iteration partial link}
\begin{array}{c}
  \Delta x(k+1)=\tilde{e}x(k+1),\\
  \vdots\\
  \Delta x(k+H_p)=\tilde{e}x(k+H_p),
\end{array}
\end{equation}
with $\tilde{e}\triangleq \left(\mbox{col}\left(e^T_{i,j}|(i,j)\in \mathcal{E}\right)\right)^T$
and $e_{i,j}\triangleq e_i-e_j$. Note that if both $(i,j)$ and $(j,i)$
are in $\mathcal{E}$, only $e_{i,j}, ~i>j$ will appear in $\tilde{e}$.
Then, substituting (\ref{eq: closed-loop system with MPC law and
  partial link}) into (\ref{eq: state difference iteration partial
  link}) yields
\begin{equation}\label{network_state_diff_H_p_steps_ahead}
\begin{array}{c}
  \Delta X(k+1)\triangleq \left[\Delta x^T(k+1),\ldots, \Delta x^T(k+H_p)\right]^T\\=\tilde{E}X(k+1)=\tilde{E}\left[\left(P_{\epsilon}+\Theta\right)^T,\ldots, \left(\left(P_{\epsilon}+\Theta\right)^{H_p}\right)^T\right]^Tx(k),
\end{array}
\end{equation}
with
$\tilde{E}=\mbox{diag}(\underbrace{\tilde{e},\ldots,\tilde{e}}_{H_p})$.

Analogous to what we have considered in the all-to-all link network
case (\ref{eq: optimization index}), the optimization index of the
network here is designed as follows:
\begin{equation}
  \label{eq: optimization indexof partial link}
\begin{array}{c}
  J(\Theta,k)=\|\Delta X(k+1)\|_Q^2+\|\Theta x(k)\|_R^2,
\end{array}
\end{equation}
where $Q$ and $R$ are positive definite symmetric weighting
matrices. Thus, it follows from
(\ref{network_state_diff_H_p_steps_ahead}) and (\ref{eq: optimization
  indexof partial link}) that
\begin{equation}
  \label{eq: optimization indexof partial link-improve}
  J(\Theta,k)=x^T(k)J_{in}(\Theta)x(k)
\end{equation}
with $J_{in}(\Theta)=\Gamma^T\tilde{E}^TQ\tilde{E}\Gamma+\Theta^TR\Theta$
and $\Gamma = \left[\left(P_{\epsilon}+\Theta\right)^T,\ldots,
  \left(\left(P_{\epsilon}+\Theta\right)^{H_p}\right)^T\right]^T$.

Consequently, control law $\Theta$ can be derived as follows
\begin{equation} \label{eq: partial link control law}
  \partial J_{in}(\Theta)/\partial \theta_{ij}|_{(i,j)\in \mathcal{E}}=0.
\end{equation}
which leads to a set of polynomial equations. To solve this, Groebner
basis methods can typically be used if the dimension of the problem to
solve is not too high. Indeed, such methods allow to find all the
solutions to a set of polynomial equations in several indeterminates,
if the solution set consists of a finite set of isolated points
\cite{co97}. Note that in the partial link case (\ref{eq: closed-loop
  system with MPC law and partial link})--(\ref{eq: partial link
  control law}), the control horizon $H_u$ is always one. Certainly,
one can extend $H_u$ by setting different values of $\Theta$ for
different future steps; however, the computational burden will thereby
increase remarkably. For this reason, multiple-step control horizons
are not preferred for partial link networks.



\subsection{Analysis}
In this section, based on {\it Lemma \ref{Th: partial link lemma}}, we
provide a necessary and sufficient condition ensuring asymptotic
convergence of the partial link MPC consensus protocol (see (\ref{eq:
  closed-loop system with MPC law and partial link})--(\ref{eq:
  condition of Theta}) and (\ref{eq: partial link control law}))
towards average-consensus.


\begin{theorem}\label{Th: partial link theorem}
  Consider a dynamical network $G_x$ whose dynamics are determined by
  the MPC protocol given in (\ref{eq: closed-loop system with MPC law
    and partial link}) and (\ref{eq: partial link control law}) and
  which satisfies the conditions (\ref{eq: not change the topology})
  and (\ref{eq: condition of Theta}). Then $\lim_{k\rightarrow
    \infty}x(k)=\textit{\textbf{1}}\bar{x}(0)$ (or average-consensus is
  reached asymptotically) if and only if either of the following two
  assumptions holds

  \begin{enumerate}
  \item[\textit{\textbf{A6}}:]
    $\rho(P_{\epsilon}+\Theta-\textbf{\textit{1}}\textbf{\textit{1}}^T/N)<1$;

  \item[\textit{\textbf{A7}}:] the matrix $P_{\epsilon}+\Theta$ has a
    simple eigenvalue at $1$ and all its other eigenvalues in the open
    unit circle.
  \end{enumerate}

\end{theorem}

\textit{Proof}:  It follows from (\ref{eq: trival eigenvector}) and (\ref{eq:
    condition of Theta}) that
  $(P_{\epsilon}+\Theta)\textit{\textbf{1}}=(P_{\epsilon}+\Theta)^T\textit{\textbf{1}}=\textit{\textbf{1}}$.
  Moreover, it can be seen from (\ref{eq: not change the topology})
  that the sparsity structure of the matrix $P_{\epsilon}+\Theta$
  corresponds to the one initially imposed by $\mathcal{E}$. Thus,
  taking into consideration of \textit{Lemma~\ref{Th: partial link
      lemma}}, the following equation holds:
  \[\lim_{k\rightarrow \infty}x(k)=\lim_{k\rightarrow
    \infty}(P_{\epsilon}+\Theta)^kx(0)=\textit{\textbf{1}}\textit{\textbf{1}}^T/Nx(0)=\textit{\textbf{1}}\bar{x}(0)
 \]
 if and only if either \textit{Assumption} \textit{A6} or \textit{A7}
 is satisfied.
$~~\hfill\blacksquare$

\begin{remark}
  It will be illustrated later by simulations that \textit{Assumption}
  \textit{A6} or \textit{A7} holds when $\epsilon$ grows to be
  much larger than $1/d_{\max}$. For partial link networks, one can
  generally compute a numerical solution (see (\ref{eq: partial link
    control law})) rather than derive the corresponding analytical
  solution, thus the analysis cannot be formulated as thoroughly as
  for the all-to-all link case. However, the rationale of the
  predictive mechanism (\ref{eq: closed-loop system with MPC law and
    partial link})--(\ref{eq: partial link control law}) is quite
  clear, i.e. the role of the routine part of the protocol
  ($P_{\epsilon}x(k)$) is to drive the state $x(k)$ to the average-consensus
  point $\textit{\textbf{1}}\bar{x}(0)$, while the role of the MPC part
  ($\Theta x(k)$) is to accelerate the consensus process by minimizing
  the future state difference between each neighboring pair in the
  network.
\end{remark}

\subsection{Case study}
In this section, we present some simulation results to compare the
convergence speeds obtained using the routine consensus protocol given
in (\ref{eq: routine control}), the FDLA protocol given in (\ref{eq:
  FDLA3}), and the proposed MPC protocol given in (\ref{eq:
  closed-loop system with MPC law and partial link})--(\ref{eq:
  condition of Theta}) and (\ref{eq: partial link control law}) for
the case of partial link networks. Moreover, through these simulation
results, one can also appreciate the difference in performances of the
proposed MPC protocol for both all-to-all link and partial link
networks.


\begin{figure}[htb]
\centering \leavevmode
\begin{tabular}{cc}
\hspace*{-0.3cm}
\resizebox{4.0cm}{!}{\includegraphics[width=11.2cm]{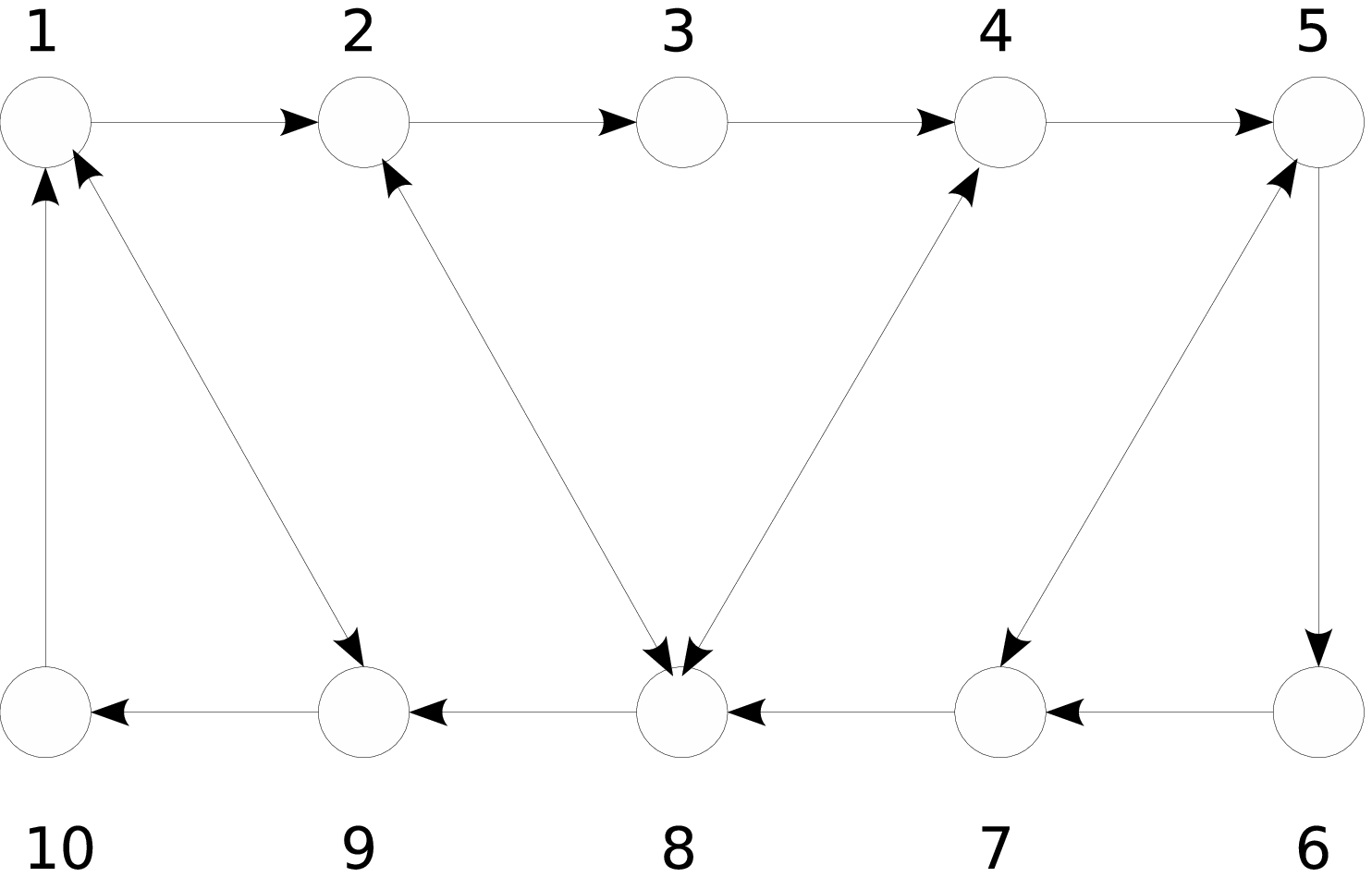}}
& \hspace*{0.3cm}
\resizebox{4.3cm}{!}{\includegraphics[width=11.2cm]{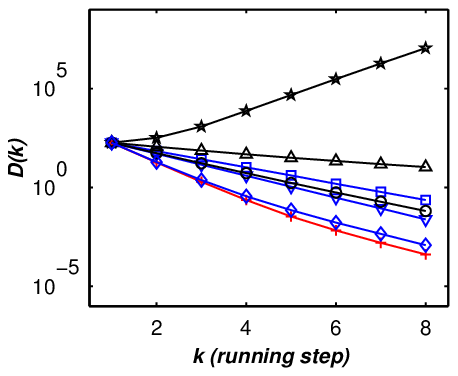}}\\
{\scriptsize (a) } &  {\scriptsize (b) }
\end{tabular}
\caption{(Color online) (a): balanced network topology; (b): time
  evolution of $D(k)=\|x(k)-\textbf{\textit{1}}\bar{x}(0)\|_2^{2}$
  when considering the MPC (blue curves), routine (black curves) and
  FDLA3 (red $+$ curve) protocols with different sampling periods
  $\epsilon$. Case~1:$\epsilon=1/3= 1/d_{\max}$, $\lozenge$: MPC,
  {\scriptsize$\bigcirc$}: routine protocol; Case~2: $\epsilon=0.01<
  1/d_{\max}$, $\bigtriangleup$: routine protocol, $\bigtriangledown$:
  MPC; Case~3: $\epsilon=1.00> 1/d_{\max}$, $\square$: MPC,
  $\bigstar$: routine protocol. In these simulations, $N=10$, $H_u=1$,
  $H_p=4$, $q=2$, and $x_i(0)~(i=1,\ldots,N)$ is selected randomly in
  $[0,15]$. $d_{\max}=\max_{i}\left(l_{ii}\right)=3$.}
\label{fig:partial convergence comparison}
\end{figure}

\begin{figure}[htb]
\centering \leavevmode
\begin{tabular}{cc}
\hspace*{-0.3cm}
\resizebox{4.0cm}{!}{\includegraphics[width=11.2cm]{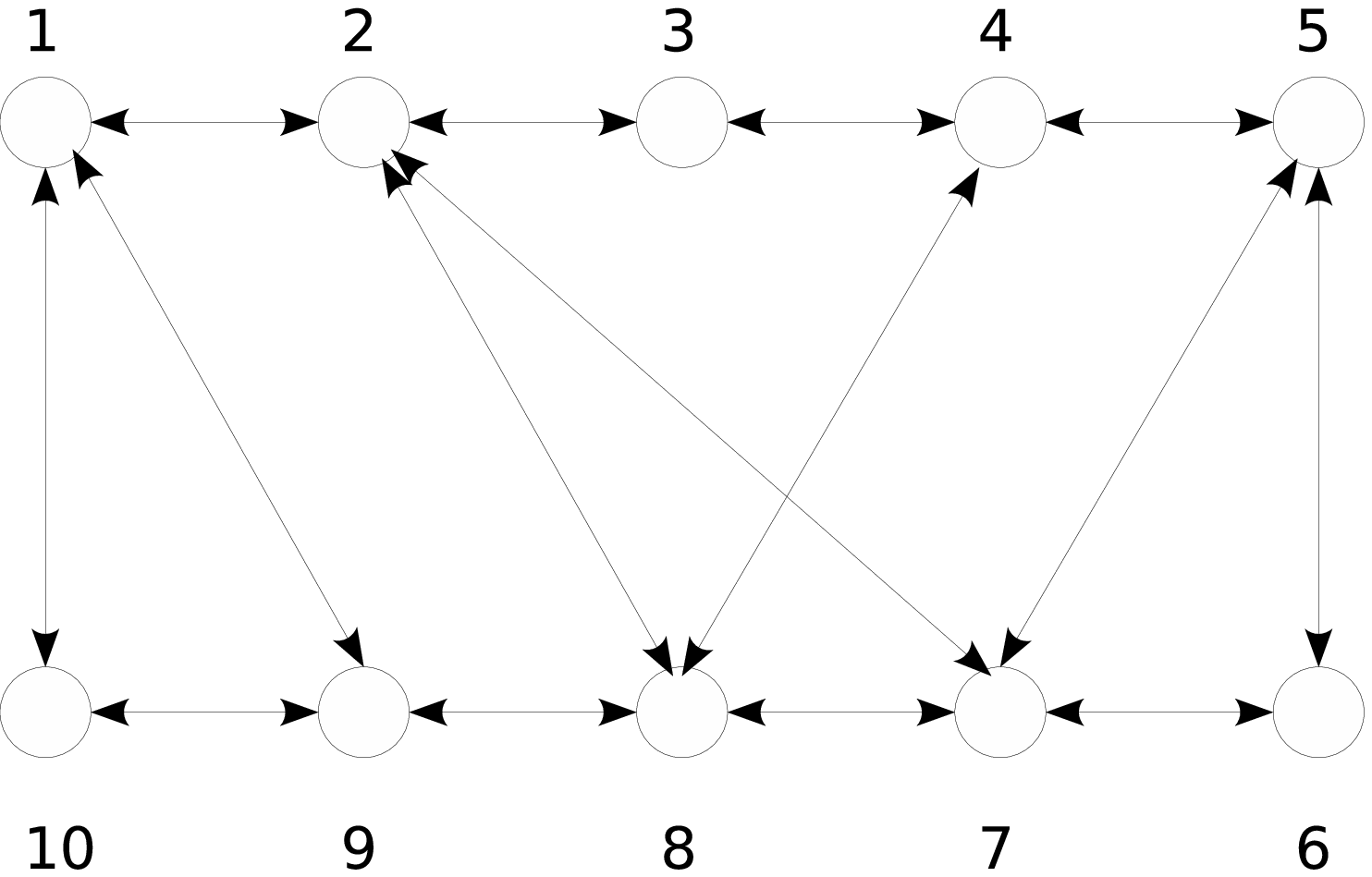}}
 &
\hspace*{0.3cm}
\resizebox{4.0cm}{!}{\includegraphics[width=11.2cm]{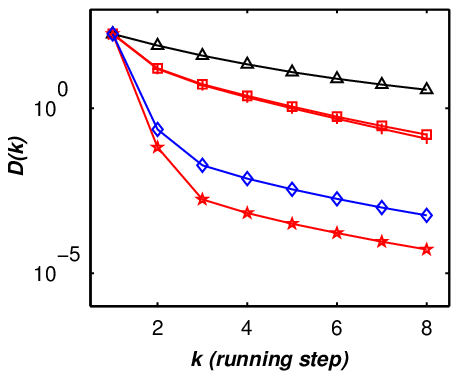}}\\
{\scriptsize (a) } &  {\scriptsize (b) }
\end{tabular}
\caption{(Color online) (a): symmetric network topology; (b): time
  evolution of $D(k)=\|x(k)-\bar{x}(0)\textbf{\textit{1}}\|_2^{2}$
  when considering the MPC (blue curves), routine (black curve) and
  FDLA3 (red curves) protocols. Here $\lozenge$: MPC;
  {$\bigtriangleup$}: \textit{routine}; {$+$}: FDLA1; $\square$:
  FDLA2; $\bigstar$: FDLA3. The MPC parameters are $N=10$, $H_u=1$,
  $H_p=4$, $q=2$, $x_i(0)~(i=1,\ldots,N)$ is selected randomly in
  $[0,15]$, $d_{\max}=\max_{i}\left(l_{ii}\right)=4$ and
  $\epsilon=0.1< 1/d_{\max}$.}
\label{fig:partial convergence comparison with FDLA}
\end{figure}

In these simulations, we have considered a $10$-node asymmetric
partial link network whose topology is given in Fig.~\ref{fig:partial
  convergence comparison}(a) (see \cite{sa04}). For this partial link
network $G=(\mathcal{V},\mathcal{E},A)$, the initial weights are fixed
as follows: if $(i,j) \in \mathcal{E},~i\neq j$, then $a_{ij}=1$;
otherwise, $a_{ij}=0$. As shown in Fig.~\ref{fig:partial convergence
  comparison}(b), three different kinds of consensus strategies,
i.e. the MPC (see (\ref{eq: closed-loop system with MPC law and
  partial link})--(\ref{eq: condition of Theta}) and (\ref{eq: partial
  link control law})), routine (see (\ref{eq: routine control})) and
FDLA3 (see (\ref{eq: FDLA3})) protocols, are compared. It can be
observed that FDLA3 still holds the highest possible convergence speed
while the addition of a predictive mechanism computed according to
(\ref{eq: closed-loop system with MPC law and partial
  link})--(\ref{eq: condition of Theta}) and (\ref{eq: partial link
  control law}) yields a drastic increase in the convergence speed
towards consensus compared with the routine protocol.  Similar phenomena
as those observed in the all-to-all link cases (see
Fig.~\ref{fig:convergence comparison}) can also be observed: for
$\epsilon\ll 1/d_{\max}$, the convergence speed is increased sharply
with this MPC protocol; for $\epsilon > 1/d_{\max}$, the routine
consensus protocol diverges while for $\epsilon$ values belonging to
the interval $(1/d_{\max},\bar{\epsilon})$ with $\bar{\epsilon}\gg
1/d_{\max}$, the MPC protocol still converges with high-speed.
Therefore, the feasible convergence range of $\epsilon$ is remarkably
expanded by the proposed MPC protocol. More interestingly, for
$\epsilon \rightarrow 1/d_{\max}$, the consensus speed of MPC
approaches the fastest possible speed yielded by FDLA3 and is thus
nearly maximized with the addition of a predictive mechanism.

To illustrate the advantages of the MPC protocol more vividly, we
present in Fig.~\ref{fig:partial convergence comparison with FDLA}(b)
a comparison of the MPC, routine, FDLA1, FDLA2 and FDLA3 protocols
(see (\ref{eq: FDLA3})) when implemented on the symmetric network
structure described in Fig.~\ref{fig:partial convergence comparison
  with FDLA}(a). In this case, the simulations were carried out with
the fixed sampling period $\epsilon=0.1$. For this partial link
network $G=(\mathcal{V},\mathcal{E},A)$, the initial weights are fixed
as follows: if $(i,j) \in\mathcal{E},~i\neq j$, then $a_{ij}=1$;
otherwise, $a_{ij}=0$. It can be observed from Fig.~\ref{fig:partial
  convergence comparison with FDLA}(b) that the consensus speeds of
MPC and FDLA3 are much higher than those of FDLA1 and
FDLA2, which is due to the larger degree of freedom of the
corresponding state matrix $W$. Compared with MPC, FDLA3's speed is
even higher, since FDLA3 yields the global optimum $W^*$ in the
fastest possible way \cite{xi04}. Moreover, it has been proven in {\it
  Theorem~\ref{Th: partial link theorem}} that consensus will be
asymptotically reached provided that \textit{Assumption} \textit{A6}
or \textit{A7} holds. Also, it can be shown by simulations that
\textit{Assumption} \textit{A6} or \textit{A7} generally holds even
when $\epsilon$ grows much larger than $ 1/d_{\max}$, and that
$(P_{\epsilon}+\Theta)^k$ typically approaches
$\textbf{\textit{1}}\textbf{\textit{1}}^T/N$ in just a few steps.

\begin{figure}[htp]
\centering \leavevmode
\begin{tabular}{cc}
\hspace*{-0.3cm}
\resizebox{4.15cm}{!}{\includegraphics[width=11.2cm]{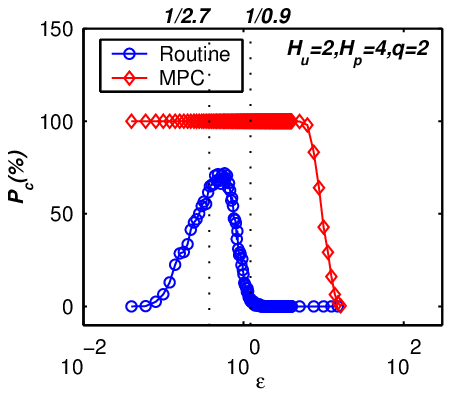}}
&
\resizebox{4.1cm}{!}{\includegraphics[width=11.2cm]{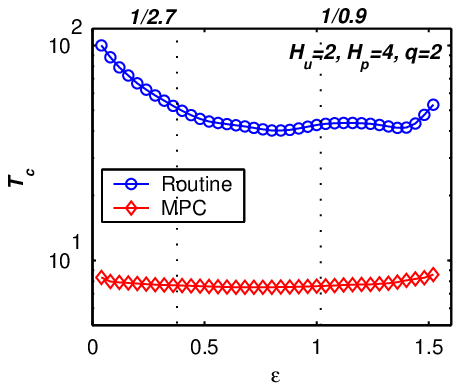}}\\
{\scriptsize (a) } &  {\scriptsize (b) }
\end{tabular}
\caption{(Color online) (a): partial link network's ultrafast-converge
  probability ($P_c(0.01)$) and (b): consensus steps
  ($T_c(0.01)$). Comparison is addressed between the MPC (red curves)
  and routine protocols (blue curves) with different sampling periods
  $\epsilon$ and $500$ independent runs for each $\epsilon$. For these
  simulations, $H_u=1$, $H_p=4$, $q=2$, the consensus threshold
  $D_c=0.01$, initial states $x_i(0),~i=1,\ldots,N$ are selected
  randomly in [0,15], and all non-zero entries $l_{ij}(i\neq j)$ of
  $L$ are selected randomly from $[-1,0]$. The corresponding values of
  $d_{\max}$ lie in the range $[0.9,2.7]$. The vertical dotted lines
  correspond to the minimum and maximum values of
  $1/d_{\max}$.} \label{fig:partial link converge probability}
\end{figure}

To compare the convergence performances of the MPC and routine
consensus protocols for the asymmetric topology described in
Fig.~\ref{fig:partial convergence comparison}, we show their
\textit{ultrafast-convergence} probabilities $P_c(0.01)$ with respect
to $\epsilon$ in Fig.~\ref{fig:partial link converge probability}(a).
As previously, a convergence run is considered ultrafast if $D_c=0.01$
can be reached within $100$ steps. It can be observed that, with the
assistance of the predictive mechanism, the ultrafast-convergence
probability is significantly enhanced for all fixed values of
$\epsilon$. Furthermore, the range of feasible convergence sampling
periods is also sharply expanded (by a factor $10$ in our simulation
results).

To further study the newly introduced predictive mechanism (\ref{eq:
  closed-loop system with MPC law and partial link})--(\ref{eq:
  partial link control law}) and to analyze its impact on the feasible
convergence range of sampling period $\epsilon$, we examine in
Fig.~\ref{fig:partial link converge probability}(b) the evolution of
the average value of the consensus steps $T_c(0.01)$ (see (\ref{eq:
  convergent step definition})) of these two strategies for the
ultrafast-consensus runs. It can be seen that, compared with the
routine consensus protocol, the use of MPC leads to a significant
increase of the consensus speed $1/T_c(0.01)$ (by a factor between $5$
and $12$ in our simulation results).

Compared with the all-to-all link network case, the performance
improvements of partial link networks are generally reduced. This
should be attributed to the fact that each node of an all-to-all
link network can use the information of all the other ones for
prediction, whereas the information flow in a partial link network
is constrained by its topology. Moreover, another interesting
phenomenon deserving notice is that, for the routine consensus
protocol, the $P_c(0.01)$ curve starts with a very low value for
small $\epsilon$, ascends to the peak (which correspond to about
$80\%$) in the left half of the $d_{\max}$ zone and finally returns
to zero very quickly within the right half of the $d_{\max}$ zone.
This phenomenon should also be attributed to the information flow
constraints imposed by the partial link topology and the fact that
the routine protocol convergence speed is maximized when
$\epsilon=1/d_{\max}$ (see FDLA2 in \cite{xi04}).

\begin{figure}[htp]
\centering \leavevmode
\begin{tabular}{cc}
\resizebox{4.36cm}{!}{\includegraphics[width=11.2cm]{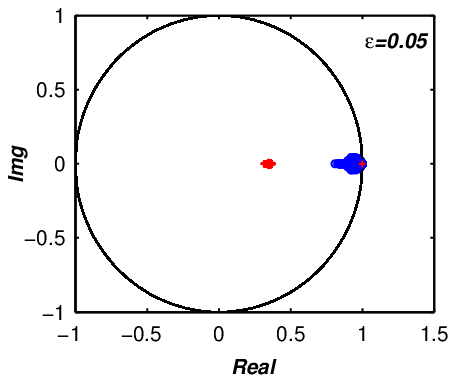}}
&
\resizebox{4.2cm}{!}{\includegraphics[width=11.2cm]{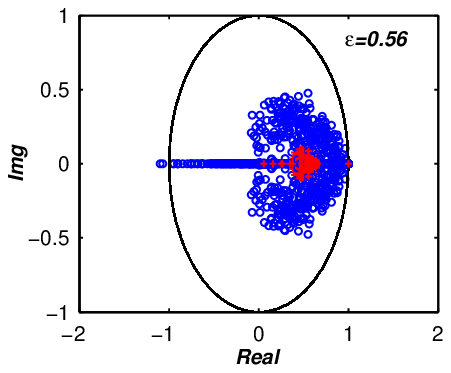}}
 \\
{\scriptsize (a) } &  {\scriptsize (b) } \\
\resizebox{4.2cm}{!}{\includegraphics[width=11.2cm]{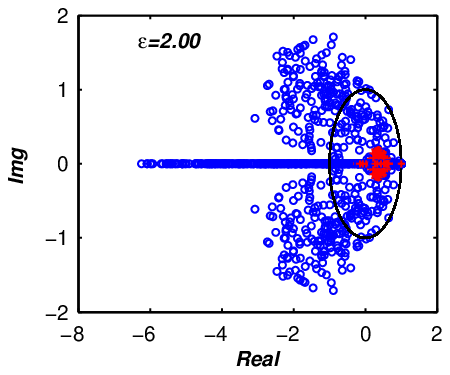}}
&
\resizebox{4.2cm}{!}{\includegraphics[width=11.2cm]{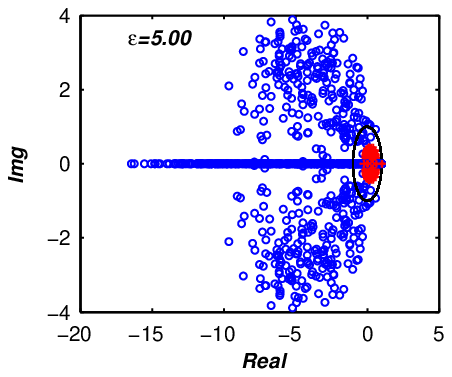}}\\
{\scriptsize (c) } &  {\scriptsize (d) }
\end{tabular}
\caption{(Color online) partial link network's eigenvalue distribution
  with respect to the sampling period $\epsilon$. Blue {\scriptsize
    {\scriptsize{\scriptsize$\bigcirc$}}} and red $+$ denote the
  eigenvalue distributions of $P_{\epsilon}$ and
  $(P_{\epsilon}+P_{MPC})$ over $100$ runs, respectively. The black
  circle represents the unit circle in the complex plane. Here,
  $H_u=1$, $H_p=4$, $q=2$. According to the network topology (see
  Fig.~\ref{fig:partial convergence comparison}(a)), each non-zero
  entry $l_{ij}~(i\neq j)$ of $L$ is selected randomly in $[-1,0]$,
  and it is calculated that $d_{\max}\in [0.9,2.7]$ in this $100$-run
  simulation.} \label{fig:eigs of partial link network}
\end{figure}

To investigate MPC's ultrafast-consensus capabilities even more
carefully, we compare the distribution of eigenvalues for the two
matrices $P_{\epsilon}$ and $P_{\epsilon}+P_{MPC}$ in
Fig.~\ref{fig:eigs of partial link network}. Similar to the all-to-all
link network case, these statistical simulations have been realized by
considering $100$ different runs for an asymmetric balanced network
with the topology shown in Fig.~\ref{fig:partial convergence
  comparison}(a). In these simulations, each non-zero entry
$l_{ij},~i\neq j$ of $L$ is selected randomly in $[-1,0]$ at each
run. For each run, we have considered $4$ different sampling period
values for $\epsilon$.  Based on these simulation results, similar
phenomena to those reported in the all-to-all link network case (see
Fig.~\ref{fig:eigenvalue distribution of MPC}) can be observed.
Compared with the eigenvalue-cluster of $P_{\epsilon}$, the one of
$P_{\epsilon}+P_{MPC}$ is always much smaller and closer to the
origin, which explains the overall higher consensus speed of the MPC
protocol. Moreover, if $\epsilon$ approaches $1/d_{\max}$
($\epsilon=0.56$, Fig.  \ref{fig:eigs of partial link network}(b)),
the eigenvalue-cluster of $P_{MPC}$ approaches that of $P_{\epsilon}$,
and the overlapping area of these two clusters increases. When
$\epsilon$ is increased beyond $1/d_{\max}$ (see Fig. \ref{fig:eigs of
  partial link network}(c) and (d)), some of the eigenvalues of
$P_{\epsilon}$ start escaping the unit circle, making the disagreement
function diverge, whereas the whole eigenvalue cluster of
$P_{\epsilon}+P_{MPC}$ remains inside the unit circle, which ensures
its convergence. Moreover, still worth mentioning is that the
eigenvalue of $P_{\epsilon}+P_{MPC}$, which corresponds to the
eigenvector $\textit{\textbf{1}}$, always remains at $1$ irrespective
of the value of $\epsilon$. This is due to the balanced constraints
(\ref{eq: condition of Theta}) imposed on the MPC matrix $\Theta$.

Finally, as shown in Fig.~\ref{fig:eigs of partial link network},
there are also obvious differences between the all-to-all link and
partial link network cases, i.e. the eigenvalue cluster of the former
is much smaller and closer to the origin than that of the latter. This
difference is also attributed to the reduced information flow imposed
by the partial link network topology (see Fig.~\ref{fig:partial
  convergence comparison}(a)). Consequently, the improvements
resulting from the use of the MPC prediction mechanism are reduced in
the partial link network scenario.


\section{Conclusion}
\label{Conclusion}

In this paper, in order to reveal the role of predictive mechanisms in
many natural bio-groups, we have added a certain kind of
predictive mechanism to the routine consensus protocol to design a
novel MPC protocol. Furthermore, we have presented mathematical
analysis as well as statistical simulation results to show the
improvement of consensus performances through the use of such a
protocol.

In particular, we have compared the routine, FDLA and MPC consensus
protocols in the general cases of symmetric and asymmetric, balanced
all-to-all link networks, and asymmetric balanced partial link
networks. The comparisons have led to the following two conclusions:
(i) the convergence speed towards consensus can be significantly
increased to approach the global optimum via the predictive mechanism,
namely, even short-term inspection of the future can produce a
significant increase in convergence speed; (ii) the sampling period
range guaranteeing convergence is increased sharply by this predictive
mechanism, giving the MPC protocol the potential to effectively save
communication energy. These advantages have been explained through mathematical analysis of the eigenvalue distributions.

To investigate the predictive mechanism more analytically, we have
provided some necessary and sufficient conditions guaranteeing asymptotic
convergence towards average-consensus for all-to-all link and partial
link networks, respectively. In particular, we have considered the
special case of balanced all-to-all link networks and showed that the
proposed MPC protocol can effectively compress the eigenvalue cluster
of the system state matrix and drive it back towards the origin of the
complex plane when the sampling period is increased beyond the
routine convergence threshold. In the special case of symmetric
all-to-all link networks, we proved that the corresponding state
matrix $P_{\epsilon}+P_{MPC}$ shares the same eigenvectors as the
matrix $P_\epsilon$.

Furthermore, to verify the generality of these conclusions, we have
also applied the proposed MPC protocol to two popular complex network
models, the Vicsek model \cite{vi95} and the Attractive/Repulsive
model \cite{ga03}. The corresponding results presented in
\cite{zh07} show that predictive protocol outperforms the routine
protocol when taking into consideration both the consensus speed and the
communication cost.

For natural science, the contribution of this work lies in its ability
to explain why networks of biological flocks/swarms/schools such as 
firefly and deep-sea fish groups do not communicate very frequently
all along but only occasionally during the whole dynamical
process. From the industrial application point of view, the value of
this work is two-fold: the consensus speed can be significantly
enhanced, and the communication energy or cost can be reduced
remarkably. All these merits are at the cost of giving the agents
the capabilities of making predictions, which can be efficiently
achieved based on currently available information. This work is just a
first attempt aiming at achieving ultrafast-consensus by injecting a
prediction mechanism into classical consensus algorithms such as the
routine consensus protocol, and we hope it will open up new avenues in
biological and industrial applications. 

Finally, we stress that the presented results rely on a centralized
control approach. In ongoing works, we are developping a decentralized
version of these MPC consensus algorithms.

\appendix
\subsection{Proof of {\it Theorem~\ref{Th:Eigenvalue
      conservation}}}\label{app1}

First, it is easy to see that the matrix $P_{U_{_{H_p N\times H_u
      N}}}$ in (\ref{eq: prediction system}) has the following
structure
\begin{equation}
\label{eq: PU}
P_U=\left[\begin{array}{llll}
    I_N           &           &           &           \\
    P_\epsilon      &   I_N       &           &           \\
    \vdots          &           &   \ddots      &           \\
    P_\epsilon^{H_u-1}  & P_\epsilon^{H_u-2}    &   \ldots      &   I_N       \\
    P_\epsilon^{H_u}    & \ldots        &   P_\epsilon^{2}  &   P_\epsilon+I_N    \\
    \vdots          &           &   \vdots      &   \vdots      \\
    P_\epsilon^{H_p-1} & \ldots & P_\epsilon^{H_p-H_u+1}
    & \begin{array}{l}
      P_\epsilon^{H_p-H_u}\\ +
      P_\epsilon^{H_p-H_u-1}\\+\ldots+I_N
                                        \end{array}
\end{array}\right].
\end{equation}
Furthermore, we have
\[P_{XE}\eta_i=\left[\lambda_{i}\eta_i^T, \ldots, \lambda_{i}^{H_p}\eta_i^T\right]^{T}.\]

The weighting matrices $Q,R$ are selected as in (\ref{eq:
  QR}). Straightforward calculations then show that
\begin{equation}\label{eq: Theorem1-1}
\begin{array}{l}
  P_{UE}^{T}QP_{XE}\eta_i=\\
  \left[\left.g_1\right._{q}^{H_p,H_u}\left(\lambda_{i}\right)\eta_i^T,\ldots,\left.g_{H_u}\right._{q}^{H_p,H_u}\left(\lambda_{i}\right)\eta_i^T\right]_{H_uN\times
    1}^{T},
\end{array}
\end{equation}
where
$\left.g_1\right._{q}^{H_{p},H_{u}}\left(\lambda_{i}\right),\ldots,\left.g_{H_u}\right._{q}^{H_{p},H_{u}}\left(\lambda_{i}\right)$
denote polynomial functions of $\lambda_{i}$ with orders determined
by $H_p$, $H_u$ and coefficients determined by $q$.

On the other hand, for $H_u$ arbitrary scalars
$\gamma_i\in\mathbb{R}~(i=1,\ldots,H_u)$, one has $P_{UE} \left[\gamma_1\eta_i^T,\ldots,\gamma_{H_u}\eta_i^T\right]^{T}=$
$\left[\left.h_1\right._{\gamma_1,\ldots,\gamma_{H_u},q}^{H_p,H_u}\left(\lambda_{i}\right)\eta_i^T,\right.\ldots,\left.\left.h_{H_u}\right._{\gamma_1,\ldots,\gamma_{H_u},q}^{H_p,H_u}\left(\lambda_{i}\right)\eta_i^T\right]_{H_uN\times
  1}^{T}$, where
$\left.h_1\right._{\gamma_1,\ldots,\gamma_{H_u},q}^{H_p,H_u}\left(\lambda_{i}\right),\ldots,\left.h_{H_u}\right._{\gamma_1,\ldots,\gamma_{H_u},q}^{H_p,H_u}\left(\lambda_{i}\right)$
are the corresponding polynomial functions with orders determined by
$H_p$, $H_u$ and coefficients determined by
$\gamma_1,\ldots,\gamma_{H_u}$ and $q$.

Since $P_{\epsilon}$ is symmetric and $Q=qI$, $R=I$, it follows that
\begin{equation}
\label{eq: Theorem1-2}
\begin{array}{c}(P_{UE}^{T}QP_{UE}+R)\left[\gamma_1\eta_i^T,\ldots,\gamma_{H_u}\eta_i^T\right]^{T}
=~~~~~~~~~~~~\\\left[\left.f_1\right._{\gamma_1,\ldots,\gamma_{H_u},q}^{H_p,H_u}\left(\lambda_{i}\right)\eta_i^T\right.
,\ldots,\left.\left.f_{H_u}\right._{\gamma_1,\ldots,\gamma_{H_u},q}^{H_p,H_u}\left(\lambda_{i}\right)\eta_i^T\right]_{H_uN\times
1}^{T},
\end{array}
\end{equation}
where
$\left.f_1\right._{\gamma_1,\ldots,\gamma_{H_u},q}^{H_p,H_u}\left(\lambda_{i}\right),\ldots,\left.f_{H_u}\right._{\gamma_1,\ldots,\gamma_{H_u},q}^{H_p,H_u}\left(\lambda_{i}\right)$
are the corresponding polynomial functions of $\lambda_i$ with
orders determined by $H_p$, $H_u$ and coefficients by
$\gamma_1,\ldots,\gamma_{H_u}$ and $q$.

For fixed values of $H_u$, $H_p$, $q$ and $\lambda_i$, it can be
verified that
$\left.g_i\right._{q}^{H_{p},H_{u}}\left(\lambda_{i}\right),~i=1,\ldots,
H_u$ corresponds to a constant scalar denoted by $g_i$, and
$\left.f_i\right._{\gamma_1,\ldots,\gamma_{H_u},q}^{H_p,H_u}\left(\lambda_{i}\right),~i=1,\ldots,
H_u,$ corresponds to a linear function of
$\gamma_1,\ldots,\gamma_{H_u}$ denoted by
$f_i\left(\gamma_1,\ldots,\gamma_{H_u}\right)$.  As a result, the
following set of $H_u$ linear equations
\begin{equation}
\label{eq: Theorem1-3}
f_i(\gamma_1,\ldots,\gamma_{H_u})=g_i,~i=1,\ldots, H_u,
\end{equation}
always possesses a unique solution
$\gamma_1=\gamma_1^*,\ldots,\gamma_{H_u}=\gamma_{H_u}^*$.  With this
particular solution, the right-hand sides of (\ref{eq: Theorem1-1})
and (\ref{eq: Theorem1-2}) are equal, i.e.
\[P_{UE}^{T}QP_{XE}\eta_i=(P_{UE}^{T}QP_{UE}+R)\left[\gamma_1^*\eta_i^T,\ldots,\gamma_{H_u}^*\eta_i^T\right]^{T}.\]
Using this last equality, the definition of $P_{MPC}$ given in
(\ref{eq: PMPC}) yields
\[\begin{array}{l}P_{MPC}\eta_i=-\left[I_N,\mathbf{0}_N,\ldots,\mathbf{0}_N\right]_{N\times H_u N}\left(P_{UE}^TQP_{UE}+R\right)^{-1}\\\cdot P_{UE}^{T}Q P_{XE}\eta_i=-\left[I_N,\mathbf{0}_N,\ldots,\mathbf{0}_N\right]\left[\gamma_1^*\eta_i^T,\ldots,\gamma_{H_u}^*\eta_i^T\right]^{T}\\~~~~~~~~~~~~~~~~~=-\gamma_1^*\eta_i.\\
\end{array}\]
Therefore, $P_{\epsilon}$ and $P_{MPC}$ share the same
eigenvector $\eta_i,~i=1,\ldots,N$ and the corresponding
eigenvalue $\nu_i$ of $P_{MPC}$ equals $-\gamma_1^*$ where
$\gamma_1^*$ is different for each fixed value of $ \lambda_i$.
This completes the proof of the property 1) in {\it Theorem~\ref{Th:Eigenvalue conservation}}.

On the other hand, recall the definition
$E\triangleq\mbox{diag}(e,\ldots,e)_{H_pN(N-1)/2 \times H_pN}$ and
take into consideration that $Q=qI$. We have
$E^TQE=diag(\underbrace{\Xi,\ldots,\Xi}_{H_p})$ with $\Xi\in
\mathbb{R}^{N\times N}$ being a symmetric matrix. Then,
$E^TQEP_X=\left[\left(\Xi P_{\epsilon}\right)^T,\ldots,\left(\Xi
    P_{\epsilon}^{H_p}\right)^T\right]^T=\left[P_{\epsilon}\Xi,\ldots,
  P_{\epsilon}^{H_p}\Xi\right]^T$ which leads to
$P_U^TE^TQEP_X=\left[h_1\left(P_{\epsilon},\Xi\right),\ldots,h_{H_u}\left(P_{\epsilon},\Xi\right)\right]^T_{H_uN\times
  N}$, where $h_i\left(P_{\epsilon},\Xi\right)~(i=1,\ldots, H_u)$ are
polynomial functions of $P_{\epsilon}$ and $\Xi$. Due to the symmetry
of $P_{\epsilon}$ and $\Xi$, $h_i(P_{\epsilon},\Xi)~(i=1,\ldots, H_u)$
are also symmetric.  Analogously, $P_{UE}^TQP_{UE}+R=\left\lbrace
  \psi_{i,j}(P_{\epsilon}) \right\rbrace,~i,j=1,\ldots,H_u$, where
$\psi_{i,j}(P_{\epsilon})$ are polynomial functions of
$P_{\epsilon}$. Due to the symmetry of $P_{\epsilon}$,
$\psi_{i,j}(P_{\epsilon}),~i,j=1,\ldots,H_u$ are also symmetric, thus
$(P_{UE}^TQP_{UE}+R)^{-1}$ contains $H_u\times H_u$ symmetric matrices
$\varphi_{i,j}(P_{\epsilon})$ and
$-\left[I_N,\mathbf{0}_N,\ldots,\mathbf{0}_N\right]_{N\times H_u
  N}(P_{UE}^TQP_{UE}+R)^{-1}=-\left[\varphi_{1,1}(P_{\epsilon}),\ldots,\varphi_{1,H_u}(P_{\epsilon})\right]$. Accordingly,
it is easy to see from (\ref{eq: PMPC}) that
$P_{MPC}=-\sum_{i=1}^{H_u}\varphi_{1,i}(P_{\epsilon})\cdot
h_i(P_{\epsilon},\Xi)$. Since $\varphi_{1,i}(P_{\epsilon})$ and
$h_i(P_{\epsilon},\Xi)$ are both symmetric, one has that $P_{MPC}$ is
also symmetric. This completes the proof of the property 2) in {\it
  Theorem~\ref{Th:Eigenvalue conservation}}.

\subsection{Proof of the second part of {\it Lemma~\ref{Th: partial
      link lemma}}}\label{app2}

\textit{Necessity}: (\ref{eq: lemma3-1}) $\Rightarrow$ \textit{A1}
and \textit{A3} (see \cite{xi04}).

Notice that $\lim_{k\rightarrow \infty} W^k$ exists if and only
if there is a non-singular matrix $T$ such that
\[\lim_{k\rightarrow
  \infty}W=\lim_{k\rightarrow \infty}T\left[\begin{array}{cc}
    I_{m}&\mathbf{0}\\\mathbf{0
    }&\mathbf{Z}\end{array}\right]T^{-1},\] where $\mathbf{Z}$ is a
convergent matrix, i.e.  $\rho(\mathbf{Z})<1$. One then has that
\begin{equation}
\label{eq: proof wk}
\begin{array}{c}\lim_{k\rightarrow
\infty}W^k=\lim_{k\rightarrow \infty}T\left[\begin{array}{cc}
I_{m}&\mathbf{0}\\\mathbf{0
}&\mathbf{Z}^k\end{array}\right]T^{-1}\\=T\left[\begin{array}{cc}
I_{m}&\mathbf{0}\\\mathbf{0}&\mathbf{0}\end{array}\right]T^{-1}=\sum_{i=1}^m\gamma_i
\zeta_i,\end{array}
\end{equation}
where $\gamma_i$ and $\zeta_i^T~(i=1,\ldots,N)$ are columns of $T$ and
rows of $T^{-1}$, respectively. Since each $\gamma_i\zeta_i^T$ is a
rank-one matrix and $\sum_{i=1}^N\gamma_i \zeta_i=TT^{-1}=I_N$ has
rank $N$, the matrix $\sum_{i=1}^m\gamma_i \zeta_i$ must have rank
$m$. Comparing (\ref{eq: lemma3-1}) and (\ref{eq: proof wk}) gives
$m=1$ and
$\gamma_1\zeta_1^T=\textbf{\textit{1}}\textbf{\textit{1}}^T/N$, which
implies that both $\gamma_1$ and $\zeta_1$ are multiples of
$\textbf{\textit{1}}$. In other words, $1$ is a simple eigenvalue of
$W$ and $\textbf{\textit{1}}$ is its associated left and also right
eigenvectors, i.e. \textit{Assumptions \textit{A1}} and \textit{A3}
hold.

\textit{Sufficiency}: \textit{A1} and \textit{A3} $\Rightarrow$
(\ref{eq: lemma3-1}).

If \textit{A1} and \textit{A3} hold, then there is a nonsingular
matrix $T$ such that \[W=T\left[\begin{array}{cc}
    1&\mathbf{0}\\\mathbf{0 }&\mathbf{Z}\end{array}\right]T^{-1},\]
where $\mathbf{Z}$ is a convergent matrix, i.e. $\rho(\mathbf{Z})<1$
(this can be derived using the Jordan canonical form \cite{st88}).
Let $\gamma_1, \ldots, \gamma_N$ be the columns of $T$ and $\zeta_1^T,
\ldots, \zeta_N^T$ be the rows of $T^{-1}$. Then, we have
\[\begin{array}{c}\lim_{k\rightarrow
    \infty}W^k=\lim_{k\rightarrow \infty}T\left[\begin{array}{cc}
      1&\mathbf{0}\\\mathbf{0
      }&\mathbf{Z}^k\end{array}\right]T^{-1}\\=T\left[\begin{array}{cc}
      1&\mathbf{0}\\\mathbf{0}&\mathbf{0}\end{array}\right]T^{-1}=\gamma_1
  \zeta_1.\end{array}\] It can be seen from \textit{Assumption A1}
and the Jordan canonical form \cite{st88} that $\gamma_1$
and $\zeta_1$ can be selected as normalized column eigenvector
$\textbf{\textit{1}}\cdot 1/\sqrt{N}$ and row eigenvector
$\textbf{\textit{1}}^T\cdot 1/\sqrt{N}$, respectively (note that
$\gamma_1$ and $\zeta_1$ are not unique, but their product is
unique), which implies (\ref{eq: lemma3-1}) immediately.

\section*{Acknowledgement}
The authors thank Prof. Guanrong Chen for intensive discussions, constructive suggestions and
scrupulous revisions of the manuscript. H. T. Zhang acknowledges the support of the National Natural Science
Foundation of China (NNSFC) under Grant No. 60704041, and the
Natural Scientific Founding Project of Huazhong (Central China)
University of Science and Technology under Grant No.  2006Q041B.
G.-B. Stan acknowledges the support of EPSRC under grant No.
EP/E02761X/1. T. Zhou acknowledges the support of NNSFC under Grant
No. 10635040.

\end{document}